\documentclass[a4paper,11pt]{article}
\pdfoutput=1 

\usepackage{jheppub} 
\usepackage[T1]{fontenc} 
\usepackage{array}
\usepackage{graphicx,rotate,multicol,caption}
\usepackage{amsfonts}
\usepackage{amsmath}
\usepackage{amssymb}
\usepackage{xcolor}
\usepackage{multirow,tabularx}
\usepackage{longtable}
\usepackage{lscape}
\usepackage{booktabs}
\usepackage{float}
\usepackage{mathrsfs}
\usepackage{hyperref}
\usepackage{pifont}
\usepackage{mciteplus}
\usepackage{caption}

\newcommand{\nn}{\nonumber}

\newcommand{\thbf}{\theta_{\mathrm{bf}}}
\newcommand{\UPMNS}{U_{\mathrm{PMNS}}}
\newcommand{\JCP}{J_{\mathrm{CP}}}
\newcommand{\ord}{\mathcal{O}}
\newcommand{\mbeta}{m_{\beta}}
\newcommand{\mbetabeta}{m_{\beta \beta}}
\newcommand{\ths}{\theta_{12}}
\newcommand{\thr}{\theta_{13}}
\newcommand{\tha}{\theta_{23}}
\newcommand{\chisquare}{\chi^2}

\newcommand{\nline}{ \displaybreak[0]\\}
\newcommand{\lag}{\mathscr{L}}
\newcommand{\mbf}{\mathbf}

\newcommand{\xmark}{\text{\ding{55}}}

\newcommand{\beq}{\begin{equation}}
\newcommand{\eeq}{\end{equation}}
\newcommand{\bea}{\begin{eqnarray}}
\newcommand{\eea}{\end{eqnarray}}

\title{\boldmath Neutrino masses and lepton mixing from $A_5$ and CP}

\author[a]{Andrea Di Iura,}
\author[a]{M.L. L\'opez-Ib\'a\~nez,}
\author[a]{Davide Meloni}

\affiliation[a]{Dipartimento di Matematica e Fisica, Universit\`a  di Roma Tre;\\
INFN, Sezione di Roma Tre,\\
Via della Vasca Navale 84, 00146 Rome, Italy}

\emailAdd{andreadiiura@gmail.com}
\emailAdd{maloi2@uv.es}
\emailAdd{davide.meloni@uniroma3.it}

\abstract{We analyse in detail the phenomenological implications for lepton masses and mixing derived by the breaking 
of the discrete symmetries $A_5 \times CP$ into the subgroups $Z_2 \times CP$ in the neutrino sector and $Z_5$ in the charged lepton sector. 
We derive accurate analytic expressions for the sum of the neutrino masses $\Sigma_i m_i$ as well as for the effective Majorana masses $m_\beta$ and 
$m_{\beta \beta}$ under different hypotheses for the flavon vevs and compare them with the exact numerical results obtained from the 
diagonalization of the neutrino mass matrix.}

\begin{document} 
\maketitle
\flushbottom
\setlength{\parindent}{0in}

%
\section{Introduction} \label{sec:intoduction}
%

The discovery of neutrino oscillations by \texttt{\large Super-Kamiokande} in $1998$ \cite{Fukuda:1998mi, Fukuda:1998ub} was the indisputable proof that neutrinos are massive and that their mass states are a non-trivial combination of flavour states; a possibility first concocted by Bruno Pontecorvo in $1957$ \cite{Pontecorvo:1957cp}.
Since then, a great experimental effort has been made in order to measure and understand the properties of these elusive particles.
From the theoretical perspective, it represents the first evidence of physics beyond the Standard Model (SM) since, with the SM ingredients, a mass term for neutrinos cannot be written due to three well-known reasons: the SM does not contemplate right-handed (sterile) neutrinos nor scalar triplets under $SU(2)_L$ nor nonrenormalizable operators. \\
Massive neutrinos introduce a new set of parameters associated with their mixing and masses.
In particular, the Pontecorvo-Maki-Nakagawa-Sakata matrix ($U_{\rm PMNS}$) (analogous to the CKM in the quark sector) \cite{Pontecorvo:1957cp, Pontecorvo:1957qd, Maki:1962mu, Pontecorvo:1967fh} can be interpreted as the transformation connecting the weak and mass neutrino eigenstates.
Under the assumption of neutrinos as Majorana (Dirac) particles, it can be parametrized in terms of three angles plus three (one) phases: the solar, atmospheric and reactor angles $\lbrace \theta_{12}, \theta_{23}, \theta_{13}\rbrace$, the Dirac phase $\delta$ and two Majorana phases $\lbrace \alpha, \beta\rbrace$.
On the other hand, neutrino experiments are not sensitive to absolute neutrino masses but only to squared mass differences. 
Many of these parameters have been measured and, contrary to what is observed in the coloured sector, neutrinos seem to exhibit a greater degeneracy in masses and admixture among generations.\\
Many approaches have been employed in the task of uncovering the origin (if any) of these peculiar patterns of masses and mixing.
One of the most appealing proposals is that of considering a flavour symmetry spontaneously broken at high energies by the non-zero vacuum expectation value (vev) of some scalar fields, commonly named {\it flavons} \cite{Froggatt:1978nt}.
Indeed, for the leptonic sector where the mixing texture seems to be specially well defined, the use of discrete groups has been particularly popular \cite{Ishimori:2010au, Altarelli:2010gt, Grimus:2011fk}.\\
Moreover, motivated by the fact that the Majorana neutrino mass matrix is always invariant under a Klein group $Z_2\times Z_2$, one may go a step further and assume that the breaking of the family symmetry ${\cal G}_\ell$ for leptons occurs in such a way that some residual subgroups remain conserved in each sector, ${\cal G}_\nu,\, {\cal G}_e \subseteq {\cal G}_\ell$.
The natural choice for neutrinos is obviously the Klein group whereas, for charged leptons, an Abelian subgroup of the initial group that distinguishes between generations is the required option \cite{Lam:2007qc}. 
This type of framework can determine all three mixing angles and the Dirac phase from symmetric principles, if the left-handed (LH) leptons transform as one of the triplet irreducible representation of the original flavour group.\\
A variation of this setup reduces the symmetry in the neutrino sector and considers the direct product of a $Z_2$ subgroup of ${\cal G}_\ell$ and CP as residual symmetry.
This choice determines all physical phases and mixing angles in terms of a single real parameter $\theta$.
In particular, the small value of the $\theta_{13}$ may be well accommodated in terms of it.
Although this formulation does not constrain lepton masses neither the free parameter $\theta$, which must be achieved in concrete models or fitted to reproduce the experimental data, testable relations between mixing angles and phases can be derived.
Also Majorana phases are predicted.
This approach has been implemented for several family symmetries, see for instance \cite{DiIura:2015kfa, Feruglio:2012cw, Holthausen:2012dk, Ding:2013bpa, Feruglio:2013hia, Ding:2013hpa, Girardi:2013sza, Rong:2016cpk, Ding:2014ssa, Ding:2014hva, Hagedorn:2014wha, Ding:2014ora, King:2014rwa, Li:2016ppt, Li:2015jxa, Ballett:2015wia, Branco:2015hea, Li:2016nap, Luhn:2013vna, Li:2013jya, Li:2014eia, Branco:2015gna, Ding:2013nsa, Hagedorn:2016lva, Hagedorn:2018gpw}. \\
Here we aim to study in detail this type of configuration for the flavour group $A_5\times {\rm CP}$. 
We will discuss explicit realizations for the neutrino mass spectrum providing relations among the mixing angles and masses.
Also predictions for the Majorana effective masses will be derived.
The paper is organised as follows: in Section \ref{sec:a5cpmix} we review the main results from \cite{DiIura:2015kfa} regarding lepton mixing and compute the structure of the neutrino masses and flavon vevs which are symmetric under the considered residual symmetry; Section \ref{sec:analytics} is devoted to obtain analytic relations for the neutrino phenomenology, based on perturbation theory; in Section \ref{sec:numerics} we perform a numerical scan to evaluate the validity of the analytical results and extract estimations for the Majorana effective masses; we conclude in Section \ref{sec:summary} summarising the most important results and providing additional remarks.

%
\section{Lepton description from ${\bf A_5}$ and CP}
\label{sec:a5cpmix}
%
The alternating group $A_5$ describes the even permutations on five elements \cite{Ishimori:2010au}.
It has $60$ elements and five conjugacy classes, thus five irreducible representations are possible: one singlet $\bf 1$, two triplets $\bf 3$ and $\bf 3'$, one tetraplet $\bf 4$ and one pentaplet $\bf 5$.
All elements in the group can be written in terms of two generators, $s$ and $t$, that satisfy:
\beq s^2= e, \hspace{1.cm} t^5=e \hspace{0.5cm} {\rm and} \hspace{0.5cm} (st)^3=e, \eeq
with $e$ the neutral element of the group.
In the $\bf 3$ representation, the generators $s$ and $t$ can be taken as\footnote{Lower case letters refer to the abstract elements of the group whilst capital letters stand for their representation in a specific basis.}
\beq \label{eqn:A5generators}
	S \:=\: \frac{1}{\sqrt{5}} \left(\begin{matrix}
			 1         & ~\sqrt{2}~ & ~\sqrt{2}~ \\
			~\sqrt{2}~ & -\varphi     &   1/\varphi   \\
			~\sqrt{2}~ & 1/\varphi     &   -\varphi
			\end{matrix} \right)
	\hspace{1.5cm}
	T \:=\: \left(\begin{matrix}
			~1~ &       0        & 0                 \\
			 0  & ~e^{i\, \Phi}~ & 0                 \\
			 0  &       0        & ~e^{4\, i\, \Phi}~
			\end{matrix} \right)
\eeq
with $\varphi=(1+\sqrt{5})/2\approx 1.62$ the Golden Ratio (GR) and $\Phi=2\pi/5$. The generators in the ${\bf 1}$, ${\bf 3'}$, ${\bf 4}$ and ${\bf 5}$ representation are collected in Appendix \ref{app:generators}.
Several subgroups are present in $A_5$: fifteen elements associated with a $Z_2$ symmetry, which additionally produces five possible Klein subgroups $Z_2 \times Z_2$, ten $Z_3$ transformations and six $Z_5$ subgroups.
Combinations of them can be considered as residual symmetries for charged leptons and neutrinos, once the initial family symmetry is spontaneously broken.\\
Here we are interested in the direct product of $A_5$ and ${\rm CP}$, which modifies the standard GR Mixing coming from pure $A_5$ \cite{Feruglio:2011qq} and introduces a continuous variable $\theta$ which parametrizes this departure.
The ${\rm CP}$ transformation corresponds to an automorphism of the flavour group \cite{Holthausen:2012dk, Chen:2014tpa, Grimus:1995zi} that in general acts non-trivially on the flavour space \cite{Grimus:1995zi, Ecker:1983hz, Ecker:1987qp, Neufeld:1987wa} and can be chosen as a unitary and symmetric matrix $X$ \cite{Feruglio:2012cw}.
Its action must be consistent with the group transformations so
\beq \label{eqn:CPcond} (X^{-1}\, A\, X)^\star \:=\: A' \eeq
must be fulfilled for $A,\, A'\in A_5$.
In particular, $A=A'$ in the case of $A_5$ so that the transformation is an isomorphism.
The condition given in Eq.~\eqref{eqn:CPcond} is invariant under a change of basis defined as
\beq \label{eqn:UnitTrans} X \longrightarrow X'=\Omega^\dagger\, X\, \Omega^\star \hspace{1.cm } A \longrightarrow A'=\Omega^\dagger\, A\, \Omega,\eeq
with $\Omega$ unitary.\\
The discussion about the combinations of residual symmetries of $A_5\times {\rm CP}$ that accommodate well the observed oscillation parameters has been performed in \cite{DiIura:2015kfa, Li:2015jxa, Ballett:2015wia}. It has been found that, for lepton {\small $SU(2)_L$}-doublets in the ${\bf 3}$ representation of {\small $A_5$}, only four possibilities reproduce the mixing angles in the allowed $3\,\sigma$ confidence region: two mixing patterns emerged from the case ${\cal G}_e=Z_5$, one from ${\cal G}_e= Z_3$ and another from ${\cal G}_e=Z_2\times Z_2$; in the neutrino sector, the direct product of $Z_2$ and ${\rm CP}$ was always assumed. In \cite{DiIura:2015kfa}, the combinations of symmetries were expressed as tuples of generators $Q$ of ${\cal G}_e$, $Z$ of $Z_2$ and $X$ of ${\rm CP}$. 
Here we are interested in the case ${\cal G}_e=Z_5$ and we consider as representative tuple
\beq \label{eqn:tuple} (Q, Z, X)=(T, T^2ST^3ST^2, X_0), \eeq
where $X_0$, in the ${\bf 3}$ representation, is given by the projector
\begin{align} \label{P23_definition}
	X_0 = P_{23} \equiv \begin{pmatrix}
			~1~ & ~0~ & ~0~ \\
			0 & 0 & 1 \\
			0 & 1 & 0
			\end{pmatrix}.
\end{align}
The form of $X_0$ in other representations is shown in Appendix \ref{app:generators}.
Notice that the generalised ${\rm CP}$ transformation defined by Eq.~\eqref{P23_definition} corresponds to a $\mu\tau$ reflection plus the canonical ${\rm CP}$ transformation, $X^c={\bf 1}$.
Applied on the neutrino sector, it invokes the permutation of a muon neutrino (antineutrino) to a tau antineutrino (neutrino) and generates a maximal atmospheric mixing, since the second and third rows of the $U_{\rm PMNS}$ must have the same absolute values.\\
The tuple in Eq.\eqref{eqn:tuple} reproduces the phenomenology of Case II in \cite{DiIura:2015kfa}, for which the best fit point (BFP) gives
\beq \label{eq:bfp} \theta_{\rm bf} \;=\; 0.175. \eeq
With the values in \cite{GONZALEZGARCIA_ET_AL}, this is one of the cases for which the $\chi^2$ analysis reflects one of the best agreement with the data for both normal and inverted ordering.
Here we aim to complete that analysis fully inspecting the neutrino mass spectrum and obtaining predictions which could be tested in present and future facilities.

\subsection{Lepton mixing and masses}
\label{subsec:review}
%
Some aspects about the lepton mixing and mass matrices can be inferred from the residual symmetries in the neutrino and charged-lepton sector \cite{Feruglio:2012cw}.
For instance, if $Q$ is a generator of ${\cal G}_e$, the charged-lepton mass matrix $m_e$ is constrained to verify 
	\beq \label{eqn:CHLmass} Q^\dagger\, m_e^\dagger m_e\, Q \:=\: m_e^\dagger m_e. \eeq
A direct consequence of Eq.~\eqref{eqn:CHLmass} is that, for $U_e$ a unitary transformation diagonalising $Q$,
	\beq U_e^\dagger\, m_e^\dagger m_e\, U_e \eeq
is also diagonal.
The selected tuple in Eq.~\eqref{eqn:tuple} takes $Q=T$ as generator of {\small ${\cal G}_e=Z_5$}. In the considered basis $T$ is diagonal, see Eq.~\eqref{eqn:A5generators}.
Therefore, {\small $U_e={\bf 1}$} and no contribution from charged leptons to the PMNS matrix is expected.
On the other hand, the light neutrino mass matrix $M_\nu$ must satisfy the conditions of invariance under $Z_2 \times {\rm CP}$
	\begin{subequations} \label{neutrino_mass_conditions}
		\begin{align}
		Z^T\, M_\nu\, Z &= M_\nu\\
		X\, M_\nu\, X &= M_\nu^\star\,,
		\end{align}
	\end{subequations}
which defines the following texture for the neutrino mass matrix:
	\begin{align} \label{Mnu_caseII}
		M_\nu = m_0 \begin{pmatrix}
          s + x + z & \dfrac{3}{2\sqrt{2}}(z + i \varphi y) & \dfrac{3}{2\sqrt{2}}(z - i \varphi y)\\
          \dfrac{3}{2\sqrt{2}}(z + i \varphi y) & \dfrac{3}{2} (x + i y) & s - \dfrac{x + z}{2} \\
          \dfrac{3}{2\sqrt{2}}(z - i \varphi y)  & s - \dfrac{x + z}{2} & \dfrac{3}{2} (x - i y) 
         \end{pmatrix}.
	\end{align}
The parameters $s, x, y$ and $z$ are dimensionless and reals and $m_0$ is the absolute neutrino mass scale.
Notice that in the limit $y \to 0$ the neutrino mass matrix is invariant under the $\mu - \tau$ symmetry.
The invariance of Eq.~\eqref{eqn:CPcond} under a change of basis as in Eq.~\eqref{eqn:UnitTrans} (now with {\small $A=Z$} since we are considering {\small ${\cal G}_\nu=Z_2\times {\rm CP}$}) ensures that $Z$ can be always taken diagonal and $X$ canonical simultaneously.
Because of that, one may prove that the mass matrix in Eq.~\eqref{Mnu_caseII} is diagonalized by \cite{Feruglio:2012cw}
	\begin{align} \label{PMNS_caseII}
		\UPMNS = \Omega_{\mathrm{II}} R_{13}(\theta) K_\nu\,,
	\end{align}
where the unitary matrix $\Omega_{\mathrm{II}}$ encondes the basis transformation related to the GR mixing
	\begin{align}
		\Omega_{\mathrm{II}} = \frac{1}{\sqrt{2}}\begin{pmatrix}
							\sqrt{2}\,\cos\phi & \sqrt{2}\,\sin\phi & \quad 0 ~\\
							\sin\phi & -\cos\phi & \quad i ~\\
							\sin\phi & -\cos\phi & \;-i ~
						\end{pmatrix},
	\end{align}
with $\sin\phi \equiv 1/\sqrt{2 + \varphi}$ and $\cos\phi \equiv \sqrt{(1 + \varphi)/(2 + \varphi)}$.
It is the matrix that satisfies {\small $X_0=\Omega_{\rm II}\, \Omega_{\rm II}^T$} and it block-diagonalises the initial neutrino mass matrix in Eq.~\eqref{Mnu_caseII}.
Then, a rotation $R_{13}(\theta)$ in the $1-3$ plane of an angle $\theta$
	\beq R_{13}(\theta) \;=\; \left(\begin{matrix}
							~ \cos\theta ~ & ~ 0 ~ & ~ \sin\theta ~ \\
							      0    & 1 & 0			\\
						   -\sin\theta & 0 & \cos\theta
						   \end{matrix} \right)
	\eeq
completely diagonalises $M_\nu$.
The size of the angle $\theta$ is entirely determined by the elements of the block-diagonalised matrix.
In our case:
\begin{align} \label{tan2theta_rel}
	\tan 2 \theta \;=\; \frac{2\sqrt{7 + 11 \varphi}\, y}{2 x(\varphi +1) \:+\: z(2\varphi + 
	1)}\,.
\end{align}
As expected, the value of $\theta$ is independent of the overall sign in neutrino mass matrix $M_\nu$. 
Finally, $K_\nu$ in Eq.~\eqref{PMNS_caseII} is a diagonal matrix with entries $\lbrace \pm 1,\, \pm i\rbrace$ needed to have positive mass spectrum.\\
Eqs.~\eqref{PMNS_caseII}-\eqref{tan2theta_rel} allow us to compute explicitly the $U_{\rm PMNS}$.
Using the standard parametrization detailed in Appendix \ref{sec:Invariant}, the mixing angles and ${\rm CP}$ phases can be inferred.
For instance, from Eq.~\eqref{mixing_angles_from_PMNS}, we observe that the atmospheric angle is fixed to be maximal, as anticipated before from considerations on the generalised ${\rm CP}$ transformation.
Similarly, the reactor angle turns out to be related to the internal angle $\theta$ through the relation
	\begin{align} \label{sin13_expression}
		\sin^2\theta_{13} = \frac{2 + \varphi}{5}\sin^2\theta\,.
	\end{align}
Therefore, Eq.~\eqref{sin13_expression} tells us that 
a small value of $\theta$ is needed in order to produce $\theta_{13} \sim 9^\circ$. 
Indeed, this is confirmed by the best fit value obtained in the $\chi^2$ analysis performed in \cite{DiIura:2015kfa}, which gives the $\theta\ll 1$ value presented in Eq.\eqref{eq:bfp}.
Additionally, Eq.~\eqref{tan2theta_rel} shows that such a small value of $\theta$ implies:
\beq \label{eqn:hierarchy} |y| \ll |x|, |z|. \eeq
Then, $y$ can be identified as a good expansion parameter and perturbation theory can be implemented in order to obtain $\sin^2\theta$ up to first order in $y^2$ as:
	\begin{align} \label{sin_theta_rel}
		\sin^2\theta = \frac{(11 \varphi +7) y^2}{\left[2 (1+\varphi) x+ (1+2 \varphi) z\right]^2} + 	\ord(y^4)\,.
	\end{align}
Eq.~\eqref{mixing_angles_from_PMNS} also indicates that the solar mixing angle $\theta_{12}$ is related to $\theta_{13}$ by the sum rule:
\begin{align} \label{s12_rel_s13}
	\sin^2\theta_{12} = \frac{\sin^2\varphi}{\cos^2\theta_{13}} \approx \frac{0.276}{\cos^2\theta_{13}}.
\end{align}
Inserting the best fit value for $\theta_{13}$ in Eq.~\eqref{s12_rel_s13}, we get $\sin^2\theta_{12} \approx 0.283$, which is within the $3\,\sigma$ allowed range, see Table \ref{tab:Maltoni_et_al}.
The Jarlskog invariant is \cite{PhysRevLett.55.1039}
	\begin{align}
		\JCP \;=\; {\rm Im} \bigg[U_{11} U^*_{13}U^*_{31}U_{33}\bigg] \;=\; -\frac{\sqrt{2 + \varphi}}
		{20}\sin2\theta\,.
	\end{align}
The Dirac phase $\delta$ can be computed from $\JCP$ as usual, see Eq.~\eqref{eqn:deltaDirac}, and is maximal in this case, $|\sin \delta| = 1$.
Since we are not imposing any constraints on $\delta$ and the quantity $\tan2\theta \simeq \sin2\theta$ up to first order in $\theta$, the sign of $\JCP$ is related to the sign of $\theta$ and hence we expect two solutions. 
The other CP invariants $I_1$ and $I_2$ are defined as 
	\bea
 		I_1 & \equiv & {\rm Im}{\bigg[U_{12} U_{12}U^*_{11}U^*_{11}\bigg]} = \sin^2\theta_{12}
 		\cos^2\theta_{12}\cos^4\theta_{13}\sin\alpha \label{eqn:I1} \\
		I_2 & \equiv & {\rm Im}{\bigg[U_{13} U_{13}U^*_{11}U^*_{11}\bigg]} = \sin^2\theta_{13}
		\cos^2\theta_{12}\cos^2\theta_{13}\sin\beta, \label{eqn:I2}
	\eea
so that the Majorana phases, $\alpha$ and $\beta$, can be extracted from Eqs.\eqref{eqn:I1} and \eqref{eqn:I2}.	
As $I_1$  and $I_2$ exactly vanish in this case, $\alpha$ and $\beta$ are zero or $\pi$.\\
The neutrino masses can be obtained from the diagonalization of $M_\nu$ as
\begin{align}
	\UPMNS^T\, M_\nu\, \UPMNS = {\rm diag}\{ \tilde{m}_1, \tilde{m}_2, \tilde{m}_3\},
\end{align}
where $\tilde{m}_j$ are the complex masses of the Majorana neutrinos. The absolute value of $\tilde{m}_j$ is indicated as $m_j$. Starting from our mass matrix we have:
\begin{align} 
	m_1 &=m_0 \bigg|s - \frac{x}{2} + z \frac{3\varphi - 2}{4} - \frac{3}{4}(\varphi - 2)\sqrt{[2 (1+\varphi) x+ 
	(1+2 \varphi) z]^2+  (28 + 44 \varphi) y^2}\bigg| \nn \\
	m_2 &=m_0  \bigg|s + x + z\bigg(1 - \frac{3}{2}\varphi \bigg) \bigg| \label{weinberg_neutrino_masses} \\
	m_3 &=m_0 \bigg| s - \frac{x}{2} + z \frac{3\varphi - 2}{4} + \frac{3}{4}(\varphi - 2)\sqrt{[2 (1+\varphi) x
	+ (1+2 \varphi) z]^2+  (28 + 44 \varphi) y^2} \bigg|. \nn
\end{align}
Expressions for the atmospheric and solar mass differences are not difficult to obtain from Eq.~\eqref{weinberg_neutrino_masses} as $\Delta m^2_{21}=m_2^2-m_1^2$ and $\Delta m^2_{3\ell}=m_2^2-m_\ell^2$
with $\ell=1$ for Normal Ordering (NO) and $\ell=2$ in the case of Inverting Ordering (IO).
However they are rather cumbersome and we omit them here.\\
Finally, since the neutrino is assumed to be a Majorana particle, it could be involved in $0\nu\beta\beta$ processes.
With our convention of the $\UPMNS$ in \eqref{PMNS_definition_general}, the effective mass is given by
	\begin{align}
		m_{\beta \beta} \equiv \bigg|\sum_j m_j U_{1j}^2 \bigg|= \cos^2\theta_{13} \Big(m_1 
		\cos^2\theta_{12} +m_2 \sin^2\theta_{12}e^{i \alpha}  \Big)+ m_3\sin^2\theta_{13} e^{i \beta},
	\end{align}
where the absolute value can be ignored when $\mbetabeta$ is expressed as a function of the oscillation parameters since we always obtain positive values.
The effective mass of the $\beta$-decay is
	\begin{align}
		m_{\beta } \equiv \sqrt{\sum_j m_j^2 |U_{1j}|^2 }= \sqrt{\cos^2\theta_{13} \Big(m_1^2 \cos^2\theta_{12} 
		+m_2^2 \sin^2\theta_{12} \Big)+ m_3^2\sin^2\theta_{13} }.
	\end{align}
From Eqs.~\eqref{sin13_expression}, \eqref{s12_rel_s13} and \eqref{weinberg_neutrino_masses}, one may note that all mixing angles and masses are invariant under the replacement $\theta \to - \theta$ and, similarly because of Eq.~\eqref{tan2theta_rel}, under the exchange $y \to -y$.
Also, they are independent of the overall sign in the mass matrix, i.e. $\{s, x,y, z \}\to -\{s, x,  y, z \}$. Therefore we expect at least two pairs of solutions $\pm \{s, x, \pm y, z \}$ for each point of the parameter space compatible with the experimental data.

\subsection{Flavon vevs} 
\label{subsec:flavon_vevs}
%
The structure of the vacuum expectation values (vev) of the neutrino flavon fields $\phi_{\nu, \bf{r}}$ in a given representation ${\bf r} \in A_5$ can be computed from the invariance relation under {\small $G_\nu=Z_2\times {\rm CP}$}. The equations for the positive eigenvalue(s) of $Z$ and $X_0$ in the representation $\bf{r}$ tell us
	\bea
		\langle \phi_{\nu, {\bf r}} \rangle & = & Z_{\bf r} \langle  \phi_{\nu,{\bf r}} \rangle 
		\label{flavon_Z}, \\
		\langle  \phi_{\nu, {\bf r}} \rangle & = & X_{0,{\bf r}} \langle  \phi_{\nu, {\bf r}} 
		\rangle^\star \label{flavon_X},
	\eea
with $Z_{\bf r}$ and $X_{0,{\bf r}}$ as in Eqs.~\eqref{eqn:genrep} and \eqref{eqn:X0rep}. Inserting them in Eqs.~\eqref{flavon_Z} and \eqref{flavon_X}, the flavon vevs are constrained to assume the general structure
\begin{subequations}\label{Flavons_Vevs}
\begin{gather}
	\langle  \phi_{\nu, {\bf 1}}  \rangle^{\rm T} = v_1 \\
	\langle  \phi_{\nu, {\bf 3}}  \rangle^{\rm T} = v \left( -\sqrt{2} \varphi^{-1},\; 1,\; 1\right) \\
	\langle  \phi_{\nu, {\bf 3'}} \rangle^{\rm T} = w \left(  \sqrt{2} \varphi,\; 1,\; 1\right) \\        
	\langle  \phi_{\nu, {\bf 4}} \rangle^{\rm T}  = \left(y_r - i y_i,\; (1 + 2\varphi) y_r - i y_i,\; 
													(1 + 2\varphi) y_r + i y_i,\; y_r + i y_i \right)
													\label{flavon_phi_nu_4} \\
	\langle  \phi_{\nu, {\bf 5}} \rangle^{\rm T} = - \left( \sqrt{\dfrac{2}{3}}(x_r + x_{r,2}),\; -x_r + i \varphi x_i,\;
													x_{r, 2} - i x_i,\; x_{r, 2} + i x_i,\; x_r + i \varphi x_i \right)
													\label{flavon_phi_nu_5}
\end{gather}
\end{subequations}
where all the coefficients are reals.\\
From Eq.~\eqref{Flavons_Vevs}, one may notice that a natural way to obtain a small value of $\theta\propto y$, as required by Eq.\eqref{eqn:hierarchy}, is considering the following two-step symmetry breaking: ${\cal G}_\ell\, \to\, {\cal G}_\nu=Z_2\,\times\, Z_2\, \times\, {\rm CP}\, \to\, Z_2\,\times\, {\rm CP}$. 
The first breaking could be due to $y_i$, $x_r$ and $x_{r,2}$ different from zero.
Then, if $y$ is proportional to one of the others vevs ($v_1$, $v$, $w$, $x_i$ or $y_r$), it will remain zero under the Klein group and CP.
If the second breaking is due to any of the other vevs, then it is at this point when $y$ takes its non-zero value. 
However, if these two breakings occur at different scales, a natural suppression between $y$ and the first vevs is expected.

\subsection{Constraints on Neutrino Masses}
\label{sec:numasses}
%
In order to get testable predictions about the neutrino phenomenology, a reduction in the the number of independent parameters is desired.
One of the simplest ways to do so is considering some of the flavon vevs equal to zero\footnote{In many situations, that is equivalent to leave out some of the flavons in a model; if not, it could be arranged working out the correspondent vacuum alignment that generates it.}.
In this way, we are left with four observables (three neutrinos masses plus one mixing angle $\theta$) and three parameters; thus, sum rules and verifiable correlations can be inferred.\\
Neutrino masses are usually generated through the so-called dimension-5 Weinberg operator and several constructions exist that reproduce it.
Under the assumption of single type of new additional particles to the SM spectrum, we are left with only three plausible realizations: type I, II and III see-saw models.
Since we aim to discuss correlations within the mass spectrum and mixing angles, all these possibilities can be reduced to two: Mechanism I, which groups the Weinberg operator and type II see-saw; and Mechanism II, consisting of type I and III see-saw scenarios.
Besides, for Mechanism II two possibilities are examined: Mechanism II-1 where the Dirac mass matrix is trivial, and Mechanism II-2 where the heavy Majorana mass matrix is trivial.
Finally, one may scrutinize all possible field asignments under the flavor symmetry and see that for Mechanism I, identifying $L\sim {\bf 3}$ or ${\bf 3'}$ leads to the same conclusions, whereas for Mechanisms II-1 and II-2 there are just two independent choices, $L,\, \nu^c\sim {\bf 3}$ and $L\sim {\bf3},~ \nu^c\sim{\bf 3'}$.
More details about this discussion can be found in Appendix \ref{App:equivalence}, where a diagram collecting all cases is depicted in Figure \ref{fig:diagram_mechanism} with the selected scenarios for the analysis highlighted with a darker background.

%
\section{Analytic Results} \label{sec:analytics}
%
In this section we discuss the main features of the mechanisms described above.
We derive analytical expressions for the mass spectrum and for the sum rule of the complex masses $\Sigma$ (more details about the latter can be found in Appendix \ref{sec:mass_sum_rule}).
Then, we investigate the mass hierarchy and discuss the specific relations among the flavon vevs able to reproduce the correct mass splittings and mixing angles within the $3\sigma$ region.
Finally, we make predictions for the Majorana phases, the sum of the neutrino masses, $m_{\beta\beta}$ and $m_\beta$.

The adopted strategy is as follows:
first, we obtain an expression for $\theta_{13}$ in terms of the flavon vevs using Eqs.~\eqref{tan2theta_rel} and \eqref{sin13_expression};
second, the neutrino masses and their mass differences are computed at the appropriate perturbative order\footnote{We identified in Eq.\eqref{eqn:hierarchy} $y$ as the appropriate expansion parameter for our analytical estimates. Note that making $\theta_{13}$ numerically close to the experimental value requires a correlation among $y$ and other parameters.} so that the hierarchy can be determined imposing {\small $\Delta m^2_{21} >0$ {\normalsize and} $\Delta m^2_{32} >0$ ($m_1<m_2<m_3$)} for NO and {\small $\Delta m^2_{21} >0$ {\normalsize and} $\Delta m^2_{31} <0 $ ($m_3<m_1<m_2$)} for IO;
once this is done, we construct the corresponding ratio $r_\ell=\Delta m^2_{21}/\Delta m^2_{3\ell}$, $\ell = 1$ ($2$) for NO (IO), and introduce an ansatz of proportionality between the two remaining parameters of $M_\nu$, i.e. $s= k x$;\footnote{Here $k$ is determined requiring $r_\ell$ and $\theta_{13}$ to be in the 3$\sigma$ range. The inferred ansatze are numerically verified. In the case of a small spread for $k$, we quote only the value that accommodates well $\theta_{13}$ and the ratio $r_\ell$.}
finally, predictions for the sum rules, phases and effective masses are drawn when a natural expansion is possible.
For our numerical evaluations we use the best fit values quoted in Table \ref{tab:Maltoni_et_al} for the mass splittings and mixing angles.

%
\subsection{Mechanism I}
\label{subsec:mechanism_I}
%
In this case the lepton doublet transforms as a triplets of $A_5$ and the effective Weinberg operators generating the neutrino masses are
\begin{align}
	\lag^{\rm eff} = y_1 \frac{\Big[[(L H)^2]_{{\bf 1}}\phi_{\nu, {\bf 1}}\Big]_{{\bf 1}}}{\Lambda^2} + y_5 \frac{\Big[[(L 
	H)^2]_{{\bf 5}}\phi_{\nu, {\bf 5}}\Big]_{{\bf 1}}}{\Lambda^2},
\end{align}

where, in agreement with the product rules given in Appendix \ref{sec:KP_A5}, only the singlet and pentaplet flavons contribute and the Yukawa couplings $y_{1,5}$ are real.
After flavour and electroweak symmetry breaking, the mass matrix for the neutrinos $M_\nu$ is given by Eq.~\eqref{Mnu_caseII} with:
\begin{align}
	s \equiv y_1 \frac{ v_1}{\Lambda}\qquad x \equiv -y_5\frac{x_{r,2}}{\Lambda} \sqrt{\frac{2}{3}} \qquad y \equiv -
	y_5\frac{x_{i} }{\Lambda} \sqrt{\frac{2}{3}}\qquad z \equiv -y_5\frac{x_{r}}{\Lambda} \sqrt{\frac{2}{3}}.
\end{align}
Here $y$ is proportional to $x_i$, the pure imaginary part of the vev in the pentaplet flavon, and the absolute mass scale is given by
$$m_0 \equiv \frac{\langle  H \rangle^2}{\Lambda}$$
with {\small $\langle H \rangle = 174\, \mathrm{GeV}$} the Higgs vacuum expectation value and $\Lambda$ the UV cutoff energy scale.
In the following we analyse the phenomenology for three independent subcases: $z=0$, $x=0$ and $s=0$.

%
\subsubsection*{Mechanism I: ${\bf z = 0}$}
\label{sec:Weinberg_operator_z_Zero}
%
From Eqs.~\eqref{tan2theta_rel} and \eqref{sin13_expression} we find that the best fit point in Eq.~\eqref{eq:bfp} implies a correlation among $y$ and $x$ given by: $y \approx \pm0.19 x$; hence $y$ is the correct expansion parameter if {\small $x\sim {\cal O}(1)$}.
The mass spectrum is obtained from Eq.~\eqref{weinberg_neutrino_masses} imposing z=0.
Expressions for the solar and atmospheric mass differences are derived from them and they reveal that only the NO is acceptable provided that {\small $x>0 \wedge s<-x  ~{\rm or}~ x <0 \wedge s>-x$}.
Expanding the mass spectrum up to ${\cal O}(y^2)$, the following sum rule can be extracted:
\begin{align} \label{mass_sum_rule_z_zero}
	\Sigma \equiv \tilde{m}_1 -  \tilde{m}_2 +  (3- \varphi)\left( \tilde{m}_3-  \tilde{m}_2  \right)\sin^2\theta_{13} 
	+ \ord\left(\sin^4\theta_{13} \right)\,.
\end{align}
The Majorana phases are both vanishing and we find that 
the linear relation $s = k\, x$ reproduces the experimental value of $r_1$ for $k \in [-10^3,-6]$.
In particular, $k=-20$ is quite a good ansatz and produces {\small $r_1\approx 1.28 \sin^2\theta_{13} + \ord(\sin^4\theta_{13})$}.
The sum of the neutrino masses is proportional to a non-trivial combination of $r_1$ and $\theta_{13}$ such that:
{\small \beq \label{sum_z_zero} \sum_j m_j \:\approx\: -\sqrt{{\Delta m_{21}^2} \frac{5 \sin^2\theta_{13}-r_1 (\varphi +2)}{r_1 \sin^2\theta_{13}}} \frac{r_1 (\varphi +1)+2\sin^2\theta_{13} (\varphi +2)}{2  \Big[ r_1 (3\varphi +1)-5 \varphi  \sin^2\theta_{13} \Big]} \:\gtrsim\: 0.155~ {\rm eV} \eeq}

Notice that, although the bound in Eq.~\eqref{sum_z_zero} is perfectly compatible with the latest \texttt{\large Planck} data\footnote{\texttt{\large Planck} limit: $\sum_j m_j \leq 0.26\ \mathrm{eV}$ $@\ 95\%$ CL}, it is above the combined limit from \texttt{Planck} $\oplus$ BAO\footnote{\texttt{\large Planck} $\oplus$ BAO limit: $\sum_j m_j \leq 0.12\ \mathrm{eV}$ $@\ 95\%$ CL} \cite{Aghanim:2018eyx}.
Analytical expressions for $m_{\beta\beta}$ and $m_\beta$ in terms of $r_1$ and $\theta_{13}$ have also been computed, but they are quite cumbersome and we only quote here the numerical intervals derived in correspondence with the allowed range for $k$ (and well confirmed by our full numerical estimates):
\begin{eqnarray}
	3.86 \times 10^{-2} \ \mathrm{eV}\lesssim~ m_\beta \approx m_{\beta\beta} ~\lesssim 6.20 \times 10^{-1}  \ \mathrm{eV}\,.
\end{eqnarray}

%
\subsubsection*{Mechanism I: ${\bf s = 0}$}
\label{sec:mechanism_I_s_zero}
%
In this case the expression of $\tan 2 \theta$ is exactly equal to that quoted in Eqs.~\eqref{tan2theta_rel}, since it does not depend on $s$, and the reactor mixing angle $\theta_{13}$ can be read from Eq.~\eqref{sin13_expression}.
At the best fit point, 
$y/z \approx \pm 1/10$ must be fulfilled in order to reproduce the best fit value of $\theta_{13}$.
Inspecting the mass spectrum, we find that the condition for reproducing NO is {\small $x<0 \wedge 0<z<-2x/\varphi$ or $x>0 \wedge -2x/\varphi<z<0$}, which implies a $z$ parameter unnaturally small; since this hierarchy does not respect the symmetry arguments stated in Section \ref{subsec:flavon_vevs}, we will not discuss it in more detail.
For IO, we obtain {\small $x<0 \wedge -2 x/\varphi<z<-4$ or $x>0 \wedge -4 x<z<-2 x/\varphi$}, which does not set a strong restriction on the magnitude of $z$.
The mass spetrum expanded up to {\small $\ord(y^2)$} fulfills the exact sum rule \cite{Ma:2005sha}:
\begin{align} \label{mass_sum_rule_s_zero}
	\Sigma = \tilde{m}_1 + \tilde{m}_2 + \tilde{m}_3 .
\end{align}
A good ansatz to get the correct value of $r_2$ turns out to be  $x= kz$ with {\small $k \approx -3/10$}.
The Majorana phases, which are independent on the perturbative expansion, are $\alpha=\pi$ and $\beta=0$ (for IO) and from the sum rule $\Sigma$ in Eq.~\eqref{mass_sum_rule_s_zero}, assuming the best fit values, a prediction for $m_3$ is obtained
\beq m_{\rm min}= m_3 = 7.64 \times 10^{-4}\ \mathrm{eV}. \eeq
Then, using the relations discussed above among $x, y$ and $z$, the sum of the neutrino masses can be expressed as
\begin{align}
	\sum_j m_j =\sqrt{-\Delta m^2_{32}}\left[2- 0.14 \sin^2\theta_{13} + \ord(\sin^4\theta_{13})\right]
	\approx 9.9\times 10^{-2} \ \mathrm{eV} \,,
\end{align}
whereas for $m_{\beta}$ and $m_{\beta \beta}$ we have
\begin{align} \label{m_betabeta_weinberg_s_zero_IO}
	m_\beta &=\sqrt{-\Delta m^2_{32}}\left[0.96+0.34 \sin^2\theta_{13} + \ord(\sin^4\theta_{13})\right]
			\approx 4.79\times 10^{-2} \ \mathrm{eV}\,,
			\nline
	m_{\beta\beta} &= \sqrt{-\Delta m^2_{32}}\left[0.41+0.03 \sin^2\theta_{13} + \ord(\sin^4\theta_{13})\right]\approx 2.01\times 	
			10^{-2}\ \mathrm{eV}\,.
\end{align}

%
\subsubsection*{Mechanism I: ${\bf x = 0}$}
\label{sec:sectxr2_zero_weinberg}
%
In this case the relation $y \approx \pm 0.16 z$ reproduces the best fit value for $\theta_{13}$, which justifies the expansion of the neutrino masses in the variable $y$.
The mass spectrum can be obtained from Eq.~\eqref{weinberg_neutrino_masses} in the limit $x\to 0$.
A closer inspection to it indicates that only IO is allowed provided that $z<0 \wedge -z/4<s<-z (\varphi+3)/4 \varphi$ or $z>0 \wedge -z(\varphi+3)/4 \varphi<s<-z/4$.
At $0$-th order, the following sum rule is satisfied
\begin{align} \label{mass_sum_rule_x_zero}
	\Sigma = \tilde{m}_1 +(\varphi +1) \tilde{m}_2 -(\varphi +2) \tilde{m}_3 + \ord(\sin^2 \theta_{13})\,.
\end{align}
The Majorana phases are both non zero, so $ \alpha = \beta =\pi $.
A good ansatz to reproduce the experimental value of $r_2$ is $s=k\,z$ with {\small $k\approx-0.3$}, which gives {\small $r_2 \approx -0.14 + 4.3\sin^2\theta_{13}$}.
Using the relations derived above and the correlation between $s$ and $z$ we obtain
\begin{align} \label{eqn:MecIx0summj}
	\sum_j m_j & = \sqrt{-\Delta m^2_{32}} \left[2.71+ 7.49\sin^2\theta_{13}+\ord(\sin^4\theta_{13})\right] \approx 1.42 \times 
				10^{-1}\ \mathrm{eV}.
\end{align}
As for the case $z=0$, the result in Eq.~\eqref{eqn:MecIx0summj} is only compatible with \texttt{\large Planck} data but not with the combined limit from \texttt{\large Planck} $\oplus$ BAO.
Finally
\begin{align}
	m_\beta &=\sqrt{-\Delta m^2_{32}}\left[1.08+2.4\sin^2\theta_{13} + \ord(\sin^4\theta_{13})  \right]\approx  5.61\times 10^{-2} 
				\ \mathrm{eV}\,, \nline
	m_{\beta \beta} &= \sqrt{-\Delta m^2_{32}}\left[0.46+0.52\sin^2\theta_{13} + \ord(\sin^4\theta_{13})  \right] 
	\approx2.32\times 10^{-2} \ \mathrm{eV}. \label{m_betabeta_weinberg_x_zero}
\end{align}

%
\subsection{Mechanism II a-1}
\label{sec:mechanism_IIa1}
%
In this section we study the first example of the see-saw mechanism where the lepton doublet and $\nu^c$ are in the same $A_5$ representation and the Dirac mass matrix is trivial, that is $M_D=m_D\, P_{23}$ with {\small $P_{23}$} given in Eq.~\eqref{P23_definition} and {\small $m_D=y_D\langle H\rangle$}.
The Majorana Lagrangian is
\begin{align}
	\lag_M = \frac{y_1}{2}\Big[(\nu^c \nu^c )_\mbf{1} \phi_{\nu, \mbf{1}}\Big]_\mbf{1}  + \frac{y_5}
			{2}\Big[(\nu^c \nu^c)_\mbf{5} 
			\phi_{\nu, \mbf{5}} \Big]_\mbf{1} + {\rm h.c.}\;.
\end{align}
It gives rise to a mass matrix with the same structure of $M_\nu$ in Eq.~\eqref{Mnu_caseII} but with elements that we tag using capital letters $\lbrace S, X, Y, Z\rbrace$ and are directly related to the flavon vevs as\footnote{Notice that it is always possible to add in $\lag_M$  a direct mass term that can be reabsorbed into $y_1 $. Thus the parameter $S$ changes as $S \to S + M/\overline{v}$, where $\bar v$ is the scale of the heavy Majorana neutrinos.}
\begin{align}
	S \equiv y_1 \frac{v_1}{\overline{v}} \qquad X \equiv -y_5\frac{x_{r,2}}{\overline{v}} \sqrt{\frac{2}{3}} \qquad Y \equiv -
	y_5\frac{x_{i} }{\overline{v}} \sqrt{\frac{2}{3}}\qquad Z \equiv -y_5\frac{x_{r}}{\overline{v}} \sqrt{\frac{2}{3}}\,.
\end{align}
The absolute scale of the mass matrix, which we choose as the scale of the heavy Majorana particles, is represented by $\overline{v}$ and defined as
\begin{align}
	\overline{v}\equiv  \max \Big\{|y_1v_1| , |y_5x_r| , |y_5x_{r,2}| , |y_5x_i|\Big\} \approx \ord(10^{13}) \ \mathrm{GeV}\,.
\end{align}
The mass matrix for the light neutrinos can be computed using the usual see-saw relation $M_\nu = - (M_M^{-1})^* m_D^2$ and it is formally equal to that in Eq.~\eqref{Mnu_caseII}, although now the parameters $\lbrace s,x,y,z\rbrace$ are defined in terms of intricate linear combinations of $\lbrace S, X, Y, Z\rbrace$ that we do not report here.
It is diagonalised by the $U_{\rm PMNS}$ in Eq.~\eqref{PMNS_caseII} with the additional condition
\beq
	\tan 2\theta = \frac{2 \sqrt{11 \varphi +7} Y  [2 S +2 X-(3 \varphi -2) Z]}{2 S (2 (\varphi +1) X+2 \varphi  Z+Z)+4 (\varphi 
	+1) X^2-4 \varphi  X Z-(5 \varphi +4) Z^2}.
\label{eqn:mecIIa1_tan2O} \eeq
The reactor angle is determined by Eq. \eqref{sin13_expression} and, as $\tan 2 \theta \propto Y$, $|Y| \ll 1$ is expected to be the correct expansion parameter. Because of Eq.~\eqref{sin13_expression}, the reactor angle is invariant under $\theta \to - \theta$, which includes $Y \to -Y$ (using the vev language $x_i \to - x_i$) by Eq.~\eqref{eqn:mecIIa1_tan2O}. The same happens for the neutrino masses, so at least two solutions $\{S, X, \pm Y,Z\}$ should generate the same phenomenology. Also the predictions for the mass spectrum and mixing angles are invariant under the replacement $\{S, X, Y,Z\} \to -\{S, X, Y,Z\}$, i.e. independent on the overall sign of the mass matrix. Hence at least two pairs of solutions for each point in the observables space are predicted in total. In the following we will distinguish three independent subcases: $Z=0,X=0$ and $S=0$.

%
\subsubsection*{Mechanism II a-1: ${\bf Z = 0}$}
\label{sec:typeI_Z_zero}
%
For this case the expressions of $\tan2\theta$ and the mass spectrum at LO are the same as in Mechanism I with $z = 0$ replacing $y \to Y$, $x \to X$ and $m_j \to m_j^{-1}$, Eqs.~\eqref{tan2theta_rel} and \eqref{weinberg_neutrino_masses}.
The value of the reactor angle for the best fit point in Eq.~\eqref{eq:bfp} together with Eqs.~\eqref{tan2theta_rel} and \eqref{sin13_expression} tells us that $Y \approx \pm0.19 X$.
Both NO and IO are allowed, although IO is obtained when $S$, $X$ and $Y$ are of the same order of magnitude, which does not respect the hierarchy among vevs imposed by Eq.~\eqref{eqn:mecIIa1_tan2O}: {\small $|Y|\ll |X|,\, |S|$}.
Thus, that possibility is not analysed in more detail.
The NO is realised when for {\small $X>0$, $X/2<S<2X$} or {\small $S>2 X$}, and when for {\small $X <0$, $S<2 X$} or {\small $2 X<S<{X}/{2}$}.
Including {\small $\ord \left(Y^2\right)$} terms, the following mass sum rule is satisfied:
\begin{align} \label{sum_rule_typeI_A_Z_zero}
	\Sigma =  \frac{1}{\tilde{m}_1}-  \frac{1}{\tilde{m}_2} + (3- \varphi)\left(    \frac{1}{\tilde{m}_3}- \frac{1}{\tilde{m}_2}  \right)\sin^2\theta_{13}  
		+ \ord\left(m^{-1}\sin^4\theta_{13} \right)\,.
\end{align}
The ansatz $S=k\,X$ with {\small $k=1$ {\normalsize and} $k=44$} reproduces the best fit values of $r_1$ and $\theta_{13}$.
For $S = X$, $m_1/m_2 \approx 1$ and $m_3/m_2 \approx 2$ at LO while the Majorana phases are $ \alpha = 0$ and $ \beta = \pi$.
The case $S \approx 44X$ corresponds to a quasi-degenerate mass spectrum and both Majorana phases vanish.
The sum of the neutrino masses in each case are
\bea
	S=X:	& \quad & \sum_j m_j = \sqrt{\Delta m^2_{31}}\left[2.31+5.19 \sin^2\theta_{13}+ 	
					  \ord(\sin^4\theta_{13}) \right]\approx 1.12 \times 10^{-1}\ \mathrm{eV} ~~ \qquad \\
	S=44X:	& \quad & \sum_j m_j \gtrsim 8.0\, \sqrt{ \Delta m^2_{31}} \approx 0.39\ \mathrm{eV}, \label{typeI_Z_zero_S_big_mass_sum}
\eea
where the second estimation in Eq.~\eqref{typeI_Z_zero_S_big_mass_sum} lies in the region excluded by the recent results from \texttt{\large Planck} 
\cite{Aghanim:2018eyx}.
The effective masses are predicted to be
\bea
	S=X:	& \quad & m_\beta = \sqrt{\Delta m^2_{31}}\left[0.58+2.80 \sin^2\theta_{13}+ 
								\ord(\sin^4\theta_{13})\right] \approx 3.16 \times 10^{-2}\ 
								\mathrm{eV} \qquad \\
	S=44X:	& \quad & m_\beta = \sqrt{\Delta m^2_{31}}\left[2.46-1.80 \sin^2\theta_{13}+ 
								\ord(\sin^4\theta_{13})\right] \approx 1.20 \times 10^{-2}\ 
								\mathrm{eV}
\eea
and
\bea
	S=X:	& \quad & m_{\beta\beta} = \sqrt{\Delta m^2_{31}}\left[0.58-0.19 \sin^2\theta_{13}+ 
					  \ord(\sin^4\theta_{13})\right]\approx 2.88 \times 10^{-2}\ \mathrm{eV}
					  \label{m_betabeta_typeI_A_Z_zero_NO} \qquad \\
	S=44X:	& \quad & m_{\beta\beta} = \sqrt{\Delta m^2_{31}}\left[2.60-1.80 \sin^2\theta_{13}+ 
					  \ord(\sin^4\theta_{13})\right]\approx 1.25 \times 10^{-1}\ \mathrm{eV} \label{eqn:S44X}\,.
\eea
For $S=44\,X$, where $m_{\beta \beta} \approx \sum_j m_j/3$, we have the lower bound $m_{\beta \beta} \gtrsim 0.12\ \mathrm{eV}$, which is also in tension with the latest limits on $0\nu\beta\beta$ decay provided by \texttt{\large GERDA}\footnote{\texttt{\large GERDA}: $m_{\beta\beta}\leq 0.12$ eV @ 90\% CL.} \cite{Agostini:2018tnm} and \texttt{\large CUORE}\footnote{\texttt{\large CUORE}: $m_{\beta\beta}\leq 0.11$ eV @ 90\% CL.} \cite{Alduino:2017ehq}.

%
\subsubsection*{Mechanism II a-1: ${\bf X = 0}$}
\label{sec:typeI_X_zero}
%
Here $\tan2\theta$ and the light neutrino masses at LO can be obtained from Mechanism I with $x=0$ and the redefinition $y \to Y$, $z \to Z$ and $m_j \to m_j^{-1}$, Eqs.~\eqref{tan2theta_rel} and \eqref{weinberg_neutrino_masses}.
The reactor mixing angle obtained from Eq. \eqref{sin13_expression} shows that $Y \approx \pm0.16 Z$ at the best fit point.
From the ratios among the neutrino masses we deduce that only the NO is allowed if {\small $Z>0 \wedge (3\varphi-1)Z/4> S > -Z/4$ or $Z<0\wedge Z<S < -Z/4$}.
The following sum rule is fulfilled:
\begin{align} \label{sum_rule_typeI_A_X_zero}
	\Sigma = \frac{1}{\tilde{m}_1} +  \frac{1+ \varphi}{\tilde{m}_2} -  \frac{\varphi +2}{\tilde{m}_3} 
			 +\ord(m^{-1}\sin^2\theta_{13})\,.
\end{align} 
The experimental value of the ratio $r_1$ is reproduced with the ansatz $S = kZ$ for $k\simeq\lbrace-1/4,\,1/3,\,2/3\rbrace$.
Accordingly, $m_1/m_2 \approx 1$ and $m_3/m_2 \approx \sqrt{5}$ is obtained for $k\simeq -1/4$, $m_1/m_2\approx 1/2$, $m_3/m_2 \approx 13/2$ for $k\simeq 1/3$ and $m_1/m_2 \approx 1/3$ and $m_3/m_2 \approx 9/2$ for $k\simeq 2/3$.
The Majorana phases are $ \alpha =\beta=  \pi$ when $k\simeq\lbrace -1/4,\,1/3\rbrace$ whereas $\alpha=\pi,~ \beta=0$ for $k\simeq2/3$.
Additional predictions can be made for the total sum of the light neutrino masses, {\small $m_\beta$} and {\small $m_{\beta \beta}$}.
We summarise them in a compact form defining
\begin{eqnarray}
\label{compact}
 \left\{\sum_j m_j,\; m_\beta,\; m_{\beta\beta}\right \} & \equiv &\sqrt{\Delta m^2_{31}} \left[a +b\,\sin^2\theta_{13} + \ord(\sin^4\theta_{13})\right]\,,
\end{eqnarray}
with the adimensional coefficients $a$ and $b$ reported in Table \ref{tab:parameter_table}.
The last column of Table \ref{tab:parameter_table} refers to the value of the observable itself ($obs$).
\begin{table}[h!]
\begin{center}
\begin{tabular}{c c  c  c  c c}
\toprule
\toprule
& & $a$  & $b$  & $obs$ (eV)\\
\midrule
& $\sum_j m_j$ & $1.23$ & $4.58$ & $6.59 \times 10^{-2}$ & \\ 
$S = Z/3$&$m_\beta$ & $0.10$ & $6.91$ &$1.26 \times 10^{-2}$ & \\
&$m_{\beta\beta}$ & $-0.01 $ & $0.93 $ & $4.49 \times 10^{-4} $ & \\
\toprule
& $\sum_j m_j$ & $2.12$ & $4.78$ & $1.11 \times 10^{-1}$ & \\ 
$S = -Z/4$&$m_\beta$ & $0.50$ & $2.82$ &$2.78 \times 10^{-2}$ & \\
&$m_{\beta\beta}$ & $0.22 $ & $-1.20 $ & $9.78 \times 10^{-3} $ & \\
\toprule
& $\sum_j m_j$ & $1.29$ & $-5.91$ & $5.73 \times 10^{-2}$ & \\ 
$S = 2Z/3$&$m_\beta$ & $0.13$ & $1.30$ &$7.74 \times 10^{-3}$ & \\
&$m_{\beta\beta}$ & $-0.01 $ & $1.16 $ & $5.63 \times 10^{-4} $ & \\
\bottomrule
\bottomrule
\end{tabular}
\caption{\small \it   Coefficients of Eq.~\eqref{compact} and numerical estimates of the observables $\sum_j m_j, m_\beta$ and $m_{\beta\beta}$.}
\label{tab:parameter_table}
\end{center}
\end{table}
 
%
\subsubsection*{Mechanism II a-1: ${\bf S = 0}$}
\label{sec:typeI_S_zero}
%
In this case $\tan2\theta$ and the light neutrino masses at LO are given by those of Mechanism I with $s=0$ substituting $x \to X$, $y \to Y$, $z \to Z$ and $m_j \to m_j^{-1}$.
The experimental value of the reactor mixing angle is reproduced for $Y \approx \pm(0.19 X + 0.16 Z)$ and both orderings are in principle allowed: NO when {\small $Z> 0 \;\wedge\; -Z/4 < X < -\varphi Z/2$ or $Z< 0 \;\wedge\; -\varphi Z/2 < X < -Z/4$} and IO when {\small $Z<0 \;\wedge\; 2 (X+Z)<3 \varphi  Z$ or $Z>0 \;\wedge\; 2 (X+Z)>3  \varphi  Z$}.
The following exact sum rule is obtained \cite{Bazzocchi:2009da}:
\beq
	\Sigma =\frac{1}{\tilde{m}_1} +  \frac{1}{\tilde{m}_2} +  \frac{1}{\tilde{m}_3}. \label{sumrule_typeI_A_S_zero} 
\eeq
Analysing the mass splittings we observe that the hierarchies that reproduce the experimental values are {\small $|X|\ll |Y|\ll |Z|$} for NO and {\small $|Z|\ll |Y|\ll |X|$} for IO, which do not correspond to any symmetry argument.
Hence, we do not discuss them further.
The Majorana phases are $\alpha=\beta=\pi$ for NO and $\alpha=0,~\beta = \pi$ for IO.
Predictions for the lightest neutrino mass can be obtained from the Majorana masses and the sum rule in Eq.~\eqref{sumrule_typeI_A_S_zero}:
\bea
	{\rm NO:} & \quad & m_{\mathrm{min}} = m_1 = 1.09\times 10^{-2} \ \mathrm{eV} \\
	{\rm IO:} & \quad & m_{\mathrm{min}} = m_3 = 2.84\times 10^{-2} \ \mathrm{eV}.
\eea
 
%
\subsection{Mechanism II a-2}
\label{subsec:mechanism_IIa2}
%
Now we explore the see-saw mechanism considering a trivial structure of the Majorana mass matrix, $M_M = M P_{23}$ with $P_{23}$ as in Eq.~\eqref{P23_definition}, and the same representation for the $SU(2)_L$ doublets and singlets, see Figure \ref{fig:diagram_mechanism}.
The lagrangian responsible for the Dirac mass matrix is
{\small \begin{align} \label{lagrangian_mass_dirac_trivial_majorana_modelE}
	\lag_D &= Y_1 (\nu^c L )_\mbf{1} H_u + y_1 \Big[(\nu^c L )_\mbf{1} \frac{\phi_{\nu, \mbf{1}}}
	{\Lambda}\Big]_\mbf{1}H_u +y_3 \Big[(\nu^c L )_\mbf{3} \frac{\phi_{\nu, \mbf{3}}}{\Lambda}\Big]_
	\mbf{1}H_u+ y_5\Big[(\nu^c L)_\mbf{5} \frac{\phi_{\nu, \mbf{5}}}{\Lambda} \Big]_\mbf{1}H_u + {\rm h.c.}
\end{align}}
where $\Lambda$ is the UV cutoff and the Yukawa couplings are real.
The form of $M_\nu$ is fixed by symmetry to be as in Eq.~\eqref{Mnu_caseII} with $\lbrace s, x, y,z\rbrace$ defined in terms of the following adimensional parameters: 
\begin{align}
	f \equiv y_1 \frac{v_1}{\Lambda} + Y_1 \qquad g \equiv y_3\frac{ v}{\Lambda} \qquad
	h_r \equiv y_5 \frac{x_r}{\Lambda} \qquad h_{r,2} \equiv y_5 \frac{ x_{r,2}}{\Lambda}\qquad h_i 
	\equiv y_5  \frac{ x_i}{\Lambda}\,.
\end{align}
As the relations among them are quite involved, we only report the expression for $y$ here:
\beq \label{eqn:mecIIa2ycond} y = \frac{1}{3 \varphi } \left[h_i  \left(2 \varphi  \left(\sqrt{6}f+h_{r,2}\right)-(\varphi +3) 
	 h_r\right)-2 \sqrt{3}g (\varphi  h_r+2 h_{r,2})\right] \,.
\eeq
The relation between the internal angle and the vevs turns out to be $\tan 2\theta \propto (g + \ord(1) h_i)$, so $g$ and $h_i$ are the relevant parameters to suppress $\theta$ and to get a small value of the reactor mixing angle $\theta_{13}$, see Eq.~\eqref{sin13_expression}. Notice also that the parameter space is independent of the sign of $\{f,g\}$ because both mixing angles and mass spectrum are invariant under the replacement $\{f,g, h_r, h_{r,2}, h_i\} \to\{-f,-g, h_r, h_{r,2}, h_i\} $.

%
\subsubsection*{Mechanism II a-2: ${\bf h_i = f = 0}$}
\label{sec:case_hi_f_zero}
%
Here we consider $f = 0$, or equivalently $v_1 = 0$ and $Y_1 = 0$, and a vanishing complex part of the vev of the ${\bf 5}$ representation, $h_i=0$.
Under these assumptions, the best fit value in Eq.~\eqref{eq:bfp} imposes $g = \pm(0.13 h_r - 0.09 h_{r,2})$, so $|g|$ is the adequate expansion parameter.
Inspecting the mass spectrum, we observe that both hierarchies are allowed: NO is obtained if $ h_{r,2}<0 \;\wedge\; 0<h_r< 2(2-3\varphi)h_{r,2}/11$ or $h_{r,2}>0 \wedge 2(2-3\varphi)h_{r,2}/11<h_r <0$ while IO is obtained if {\small $h_{r,2}<0 \wedge {(2-2 \varphi ) h_{r,2}}<h_r<-4 h_{r,2} $} or {\small $h_{r,2}>0 \wedge -4 h_{r,2}<h_r<{(2-2 \varphi )  h_{r,2}}$}.
The masses satisfy the sum rule
\begin{align} \label{sum_rule_typeI_E_hi_f_zero}
	\Sigma = ( \tilde{m}_1 +  \tilde{m}_2  -\tilde{m}_3 )^2 - 4 \tilde{m}_1 \tilde{m}_2 + 
	\ord(\sin^2\theta_{13})\,,
\end{align}
where the undisplayed coefficient of order $\sin^2\theta_{13}$ is proportional to $h_r + 4 h_{r,2} \propto\sqrt{m_3/m_0}$ and thus the sum rule is presumed to work better in the case of IO.
The Majorana phases are fixed and vanishing.
A natural suppression of the solar squared mass difference (and then of $r_\ell$) is obtained for $h_r \approx 0$ (NO) and for $h_r \approx -4 h_{r,2}$ (IO).
Considering these two limits, the reactor mixing angle is approximately given by
\bea
	\sin^2\theta_{13} \approx 3 g^2/h_{r,2}^2 & \quad {\rm if}\; & h_r = 0 \quad\Longrightarrow\quad 
	g\approx \pm h_{r,2}/10\\ 
	\sin^2\theta_{13} \approx \frac{1}{15} g^2/h_{r,2}^2 & \quad {\rm if}\, & h_r=-4 h_{r,2} \quad
	\Longrightarrow\quad g\approx \pm h_{r,2}/2.
\eea
In the first case, $m_1/m_2 \approx 1$ and $m_3/m_2 \approx 4$ whereas for $h_r=-4 h_{r,2}$, $m_1/m_2 \approx 1$, $m_3/m_2 \approx 0$.
The predictions for the total sum of masses are
{\small \bea
	\sum_j m_j & = & \sqrt{\Delta m^2_{31}} \left[1.55 -1.86  \sin^2\theta_{13} + 
					  \ord(\sin^4\theta_{13})\right]\approx 7.48 \times 10^{-2}\ \mathrm{eV}, \quad 
					  h_r = 0 \\
	\nn \\
	\sum_j m_j & = & \sqrt{-\Delta m^2_{32}} \left[2+2.76\sin^2\theta_{13} + \ord(\sin^4\theta_{13})
				 \right]\approx 1.02 \times 10^{-1}\ \mathrm{eV}, \quad h_r = -4h_{r,2}. \quad 
				 \label{eqn:SumMecIIa2hif}
\eea}
Finally, we obtain
{\small \bea
	m_\beta & = & \sqrt{\Delta m^2_{31}} \left[0.26 +0.99  \sin^2\theta_{13} + \ord(\sin^4\theta_{13})
			  \right]\approx 1.39 \times 10^{-2}\ \mathrm{eV}, \quad h_r = 0 \\
	\nn \\
	m_\beta & = & \sqrt{-\Delta m^2_{32}} \left[1+ 1.50 \sin^2\theta_{13} + \ord(\sin^4\theta_{13})
				  \right]\approx 5.11 \times 10^{-2}\ \mathrm{eV} \quad h_r = -4h_{r,2} 
				  \label{eqn:SumMecIIa2mbeta}
\eea}
and
{\small \bea
	m_{\beta \beta} & = & \sqrt{\Delta m^2_{31}} \left[0.26 -0.17 \sin^2\theta_{13} + 
					  \ord(\sin^4\theta_{13})\right]\approx1.26 \times 10^{-2}\ \mathrm{eV}, \quad h_r 
					  = 0 \\
	\nn \\				  
	m_{\beta \beta} & = & \sqrt{-\Delta m^2_{32}} \left[1+\sin^2\theta_{13} + \ord(\sin^4\theta_{13})
					  \right]\approx5.06 \times 10^{-2}\ \mathrm{eV}, \quad h_r = -4h_{r,2}.
					  \label{typeI_E_m_betabeta_hi_f_zero_IO}
\eea}

%
\subsubsection*{Mechanism II a-2: ${\bf g = f = 0}$}
\label{sec:case_g_f_zero}
%
In this case we remove the contributions from the singlet and the triplet and the only flavon acting is in the ${\bf 5}$ representation.
Then, the parameter $h_i$ must be the smallest one in order to obtain a compatible value of the reactor mixing angle.
Here the LO mass matrix $M_\nu$ is the same as in the previous case with $h_i = f= 0$ and therefore mass orderings, the  sum rule $\Sigma$ and the Majorana phases can be read from the equations above.
We emphasize here that the sum rule defined in Eq.~\eqref{sum_rule_typeI_E_hi_f_zero} is exact (no  {\small $\ord(m^2\sin^2\theta_{13})$} are needed) and can be used to find the lightest neutrino mass as
\bea
	{\rm NO:} & \quad & m_{\mathrm{min}} \equiv m_1 = 1.13 \times 10^{-2} \ \mathrm{eV}, \\
	{\rm IO:} & \quad & m_{\mathrm{min}} \equiv m_3 = 2.97 \times 10^{-6} \ \mathrm{eV}.
\eea
As before, $m_3/m_0 \propto (h_r + 4h_{r,2})^2$ and $\Delta m^2_{21}/m_0^2 \propto h_r (h_r + 4 h_{r,2})$ so $h_r \approx 0 $ for NO and $h_r \approx - 4 h_{r,2}$ for IO.
Introducing these relations into Eq.~\eqref{sin13_expression},
\bea
	\sin^2\theta_{13} \approx \frac{1 + \varphi}{4}\frac{h_i^2}{h_{r,2}^2} & \quad {\rm if}\; & h_r = 0 \quad
	\Longrightarrow\quad h_i \approx \pm h_{r,2}/5\\ 
	\sin^2\theta_{13} \approx \frac{1 + \varphi}{20}\frac{h_i^2}{h_{r,2}^2} & \quad {\rm if}\, & h_r=-4 h_{r,2} 
	\quad \Longrightarrow\quad h_i \approx \pm 2h_{r,2}/5.
\eea
In the first case the spectrum is  $m_1/m_2 \approx 1$ and $m_3/m_2 \approx 4$ whereas for IO we predict $m_1/m_2 \approx1$ and $m_3/m_2 \approx 0$.
The total sum of neutrino masses and effective mass are the same as before for IO, Eqs.~\eqref{eqn:SumMecIIa2hif}, \eqref{eqn:SumMecIIa2mbeta} and \eqref{typeI_E_m_betabeta_hi_f_zero_IO}, while for NO we obtain
\begin{align}
	\sum_j m_j &=  \sqrt{\Delta m^2_{31}}\left[ 1.55 -0.43 \sin^2\theta_{13} + \ord(\sin^4\theta_{13})\right]
	\approx 7.73 \times 10^{-2}\ \mathrm{eV} \nline
	m_\beta &=  \sqrt{\Delta m^2_{31}}\left[ 0.26 + 2.49 \sin^2\theta_{13} + \ord(\sin^4\theta_{13})\right]
	\approx 1.55 \times 10^{-2}\ \mathrm{eV} \nline
	m_{\beta \beta} &=  \sqrt{\Delta m^2_{31}}\left[ 0.26 + 1.32 \sin^2\theta_{13} + \ord(\sin^4\theta_{13})
	\right]\approx1.42 \times 10^{-2}\ \mathrm{eV}\,. \label{typeI_E_m_betabeta_g_f_zero_NO}
\end{align}

%
\subsubsection*{Mechanism II a-2: ${\bf h_i = h_r = 0}$}
\label{sec:case_hi_hr_zero}
%
This case is similar to the Weinberg operator with $z = 0$ discussed in Section \ref{sec:Weinberg_operator_z_Zero}, indeed the $z$ mass matrix parameter is {\small $z = \ord(g^2)$} when $h_r = 0$.
The reactor mixing angle imposes $g \approx \mp(0.22 f + 0.09 h_{r,2})$ in the best fit point, so {\small $|g| \ll |f|, |h_{r,2}|$} and $m_1 = m_2$ in the LO approximation.
In this scenario only NO is allowed, provided that {\small $h_{r,2}<0 \wedge \sqrt{{2}/{3}} h_{r,2}<f<-{h_{r,2}}/{\sqrt{6}}$} or {\small $h_{r,2}>0 \wedge -{h_{r,2}}/{\sqrt{6}}<f<\sqrt{{2}/{3}} h_{r,2}$}.
To obtain a sum rule, we need to include terms up to {\small $\ord(\sin^4\theta_{13})$}, since the difference $(m_2 - m_1)/m_0 \propto \sin^2\theta_{13}$:
\bea \label{sum_rule_typeI_E_hihr_zero}
	\Sigma & = &\left(\tilde{m}_1 - \tilde{m}_2 \right)^2 + 2(3 -\varphi) \left(\tilde{m}_1 - 3 
			  \tilde{m}_2 \right) \left(\tilde{m}_1 -\tilde{m}_2 \right)\sin^2\theta_{13}\\
		   & + & 20(\varphi -2)  \tilde{m}_2  \left(\tilde{m}_1 - 2 \tilde{m}_2 +  
			  \tilde{m}_3\right)\sin^4\theta_{13} + \ord(m^2\sin^6\theta_{13})\,.
\eea
The Majorana phases are both vanishing.
As in Mechanism I $z=0$, a large spread for $k$ is obtained when the ansatz $f=k\,h_{r,2}$ is assumed: $k$ may go from $k = -1/\sqrt{6}$ to $k \approx -7/22$.
A particularly good value that accomodates well the reactor angle and the mass ratio is $f \approx -7/20\, h_{r,2}$, for which
\begin{align}
	m_\beta &\approx m_{\beta\beta} \approx \sqrt{\Delta m^2_{21}}\left[\frac{1.35 }{\sin^2\theta_{13}}+ \ord(\sin
	\theta_{13})\right] \approx 7.88 \times 10^{-2}\ \mathrm{eV}\,.
\end{align}
Given the spread in $k$, also lower bounds can be derived
\beq \label{eqn:mecIIa2hihrsummj}
	\mbeta (\mbetabeta) \gtrsim 5.49 \times 10^{-2}\ \mathrm{eV} \quad{\rm and}\quad \sum_j m_j \gtrsim 0.19\ \mathrm{eV}. \eeq
Notice that the limit for the sum of the neutrino masses in Eq.~\eqref{eqn:mecIIa2hihrsummj} is excluded by the latest results from \texttt{\large Planck} $\oplus$ BAO \cite{Aghanim:2018eyx}.

%
\subsubsection*{Mechanism II a-2: ${\bf g = h_r = 0}$}
\label{sec:case_g_hr_zero}
%
In this case Eq.~\eqref{eq:bfp} and the relation $h_i = \pm 0.19 h_{r,2}$ give the correct value for $\theta_{13}$; thus, we can expand the observables as a series in the parameter $h_i$.
At LO the mass spectrum, the sum rule for the neutrino complex masses and the Majorana phases are the same as those obtained in Section \ref{sec:case_hi_hr_zero} for $h_i = h_r = 0$.
However NLO corrections introduce some differences in the value of the solar mass splitting.
We find out that in this scenario only NO is allowed when {\small $h_{r,2}<0 \wedge f<\sqrt{{2}/{3}} h_{r,2}$} or {\small $h_{r,2}>0 \wedge f>\sqrt{{2}/{3}} h_{r,2}$}.\\
The linear correlation $f = k \,h_{r,2}$ tell us that $k$ living in the region {\small $k \gtrsim 170$} reproduces the experimental value of $r_1$ in the 3$\sigma$ confidence region.
The limit $k \to \infty$ (which corresponds to {\small $\ord(10^3)$} in our following scan) gives rise to a degenerate mass spectrum at LO, {\small $m_j/m_0 = f^2 +\ord(h_i^2)$}, whereas the value $k \approx 20$ is a quite good ansatz since we get $r_1 \approx 1.14\sin^2\theta_{13}$.\\
As for the case of $h_i = h_r = 0$, the relation for the sum of the neutrino masses as a function of $r_1, \Delta m^2_{21}$ and $\theta_{13}$ is rather cumbersome so we only report the following lower bound:
\beq \label{eqn:summjmecIIa2ghr} \sum_j m_j \gtrsim 0.19\ \mathrm{eV}. \eeq
The result in Eq.~\eqref{eqn:summjmecIIa2ghr} is compatible with the latest cosmological results from \texttt{\large Planck} but not with the combined one from \texttt{\large Planck} $\oplus$ BAO \cite{Aghanim:2018eyx}.
The effective masses $m_\beta$ and $m_{\beta\beta}$ can be easily evaluated  for $k \approx 20$ as
\begin{align}
	m_\beta &= \sqrt{\Delta m^2_{31}}\left[1.27 -1.70 \sin^2\theta_{13} +\ord(\sin^4\theta_{13})\right] \approx 
	6.12 \times 10^{-2}\ \mathrm{eV} \\
	m_{\beta\beta} &=\sqrt{\Delta m^2_{31}}\left[1.27 -1.75\sin^2\theta_{13} +\ord(\sin^4\theta_{13})\right]
	\approx 6.1 3\times 10^{-2}\ \mathrm{eV}\,. \label{TypeI_E_m_betabeta_g_hr_zero}
\end{align}  
Due to the spread in $k$ we get the lower bounds
\beq 5.53 \times 10^{-2}\ \mathrm{eV} \lesssim \mbeta \approx \mbetabeta \lesssim 4.85 \times 10^{-1}\ \mathrm{eV} \eeq

%
\subsubsection*{Mechanism II a-2: ${\bf h_i = h_{r,2} = 0}$}
\label{sec:case_hi_hr2_zero}
%
In the vev language $h_i = h_{r,2} = 0$ corresponds to $x_i = x_{r,2} = 0$.
The experimental value of $\theta_{13}$ is reproduced for $g \approx \pm( 0.13 h_r-0.22 f )$, which satisfies our criteria discussed after Eq.~\eqref{eqn:mecIIa2ycond}.
The IO is the only hierarchy allowed in this case when {\small $f<0 \wedge 2 \sqrt{6} f<h_r< 2 \sqrt{6}  f/(3   \varphi -2)$} or {\small $f>0 \wedge 2 \sqrt{6}  f/(3   \varphi -2)<h_r<2 \sqrt{6} f$}.
In the limit $|g| \ll |f|, |h_{r}|$, the following sum rule arises:
\begin{align} \label{sumrule_typeI_E_hi_hr2_zero}
	\Sigma = \left[  \tilde{m}_1 + (3 \varphi +2) \tilde{m}_2 -5 (\varphi +1) \tilde{m}_3\right]^2 -4 
			(3 \varphi +2) \tilde{m}_1 \tilde{m}_2 + \ord(m^2\sin^2\theta_{13})\,.
\end{align}
Since the coefficients in front of the masses are numbers of {\small $\ord(10)$}, we expect large NLO corrections. 
The Majorana phases are both vanishing.
In order to suppress the ratio $r_2$ one may make the solar mass difference almost vanishing, that is to impose $h_r \approx 2\sqrt{6}f$.
In fact, it can be checked that this ansatz is only compatible with IO, where the neutrino masses are {\small $m_1 = m_2 = 45\, m_0\, f^2 +\ord(g^2)$} and {\small $m_3 = 9\, m_0\, f^2+\ord(g^2)$}, and the reactor angle turns out to be {\small $\sin^2\theta_{13}\approx 0.15 {g^2}/{f^2}$}.
Simple predictions for $\sum_j m_j$, $m_\beta$ and $m_{\beta \beta}$ are possible,
\begin{align}
	\sum_j m_j &=\sqrt{-\Delta m^2_{32}}\left[2.25 -0.22 \sin^2\theta_{13} + \ord(\sin^4\theta_{13})
	\right]\approx 1.11 \times 10^{-1}\ \mathrm{eV}\,
	\nline
	m_\beta &= \sqrt{-\Delta m^2_{32}}\left[1.02 + 0.49\sin^2\theta_{13} + \ord(\sin^4\theta_{13})
	\right]\approx 5.10 \times 10^{-2}\ \mathrm{eV}
\end{align}
and
\begin{align} \label{typeI_E_m_betabeta_hi_hr2_zero_IO}
	m_{\beta \beta} &= \sqrt{-\Delta m^2_{32}}\left[1.02 - 0.84\sin^2\theta_{13} + 
	\ord(\sin^4\theta_{13})\right]\approx 4.96 \times 10^{-2}\ \mathrm{eV}\,.
\end{align}

%
\subsubsection*{Mechanism II a-2: ${\bf g = h_{r,2} = 0}$}
\label{sec:case_g_hr2_zero}
%
In this case the correlation needed to reproduce the reactor angle is given by $h_i = \pm 0.155 h_{r}$, so $h_i$ is the adequate expansion parameter.
The LO expressions for the neutrino spectrum are the same as in Section \ref{sec:case_hi_hr2_zero} with $h_i = h_{r,2} = 0$, thus we do not report here the predictions for the mass ordering, $\Sigma$ and the Majorana phases.
As discussed previously, only IO is allowed and we predict $m_1/m_2 \approx 1$ and $m_3/m_2 \approx 1/5$.
The sum of the neutrino masses $\sum_j m_j$ and the parameters $m_\beta$ and $m_{\beta\beta}$ are
\begin{align}
	\sum_j m_j &= \sqrt{-\Delta m^2_{32}}\left[ 2.25 + 6.75 \sin^2\theta_{13} + \ord(\sin^4\theta_{13})\right]
	\approx 1.18 \times 10^{-1}\ \mathrm{eV}\,, \nline
	m_\beta &=  \sqrt{-\Delta m^2_{32}}\left[ 1.02 + 2.84 \sin^2\theta_{13} + \ord(\sin^4\theta_{13})\right]
	\approx 5.36 \times 10^{-2}\ \mathrm{eV}
\end{align}
and
\begin{align} \label{typeI_E_m_betabeta_g_hr2_zero_IO}
	m_{\beta \beta} &= \sqrt{-\Delta m^2_{32}}\left[ 1.02 + 2.52 \sin^2\theta_{13} + \ord(\sin^4\theta_{13})
	\right]\approx 5.32 \times 10^{-2}\ \mathrm{eV}.
\end{align}

%
\subsection{Mechanism II c-2}
\label{subsec:mechanism_IIc}
%
In this realization we consider a Type I see-saw mechanism with the right-handed neutrinos and the lepton $SU(2)_L$-doublet transforming in different triplet representations of $A_5$.
The lagrangian responsible for the Dirac mass is
\begin{align} \label{lagrangian_mass_dirac_trivial_majorana}
	\lag_D = y_4 \Big[(\nu^c L )_\mbf{4} \frac{\phi_{\nu, \mbf{4}}}{\Lambda}\Big]_\mbf{1}H_d  + y_5\Big[(\nu^c 
	L)_\mbf{5} \frac{\phi_{\nu, \mbf{5}}}{\Lambda} \Big]_\mbf{1}H_d + {\rm h.c.},
\end{align}
where $\Lambda$ is the UV cutoff scale.
The heavy Majorana mass matrix is trivial, so the three right-handed neutrinos are exactly degenerate.
As before, $M_\nu$ is given by Eq.~\eqref{Mnu_caseII} where the $\lbrace s,x,y,z\rbrace$ parameters are non-trivial combinations of the adimensional parameters
\begin{align} \label{eqn:mecIIc2AdimParam}
	f_r \equiv y_4 \frac{y_r}{\Lambda} \qquad f_i \equiv y_4 \frac{y_i}{\Lambda} \qquad
	h_r \equiv y_5 \frac{x_r}{\Lambda}\qquad h_{r,2} \equiv y_5 \frac{x_{r,2}}{\Lambda} \qquad h_i \equiv y_5 
	\frac{x_i}{\Lambda}. 
\end{align}
The expressions are rather complicated and we only report the one associated with $y$ here
{\small \begin{align} \label{mass_parameters_typeI_modelB}
	y &= \frac{4}{9} \bigg[ h_i \left(2 \varphi  h_{r,2}-(\varphi -2) h_r\right) +    \sqrt{3} h_i f_i+ f_r   
	\left(\left(\sqrt{3}+\sqrt{15}\right) h_r-\sqrt{3} h_{r,2}\right)+  3 (\varphi +2) f_r f_i\bigg].
\end{align}}
The angle $\theta$ is related to the vevs by the relation $\tan 2\theta \propto (h_i + \ord(1) f_r)$, which means that we would need $h_i$ and $f_r$ ($ \leftrightarrow x_i$ and $y_r$, in the vev language) small compared to the other vevs to reproduce the reactor angle.
Notice that the case $h_i = 0$ is equivalent to $\phi_{\nu, {\bf 5}}$ invariant under $Z_2 \times Z_2 \times {\rm CP}$, while $f_r= 0$ corresponds to $\phi_{\nu, {\bf 4}}$ invariant under $Z_2 \times Z_2 \times {\rm CP}$.
For all the cases discussed in this section, the Majorana phases are vanishing.

%
\subsubsection*{Mechanism II c-2: ${\bf h_i = f_i = 0}$}
\label{sec:mechanismII_hi_fi_zero}
%
In this case we observe that the only symmetry-motivated possibility to get a small $\theta_{13}$ is to have a small $f_r$.
The masses of the light neutrinos are obtained from Eq.~\eqref{weinberg_neutrino_masses} with the conditions $h_i = f_i = 0$.
The solar mass squared difference is proportional to {\small $\left(h_r^2-h_{r,2}^2\right)$} and thus a small value of the solar splitting can be achieved for {\small $h_r \approx \pm h_{r,2}$}.
In this case the parameter $z$ in the neutrino mass $M_\nu$ is such that $z= \ord(f_r^2) = \ord(y_r^2/\Lambda^2)$, a situation rather similar to the Weinberg operator with $z = 0$ discussed in Section \ref{sec:Weinberg_operator_z_Zero}. \\
Both orderings are allowed: NO if {\small $h_r<0 \wedge h_r(\varphi-2) < h_{r,2} < -h_r$} or {\small $h_r>0 \wedge -h_r < h_{r,2} < h_r(\varphi-2)$} and  IO if {\small $h_r<0 \wedge h_r < h_{r,2} < h_r/(2 + 3\varphi)$} or {\small $h_r>0 \wedge h_r/(2 + 3\varphi) < h_{r,2} < h_r$}.
The sum rule is the same as in Eq.~\eqref{sum_rule_typeI_E_hi_f_zero} and we have verified that the  coefficient in front of the $\sin^2\theta_{13}$ term is proportional to {\small $(h_r - h_{r,2}) \approx \sqrt{m_3/m_0}$}, thus we expect that the sum rule works better in the case of IO.
Using the fact that $h_r \approx \pm h_{r,2}$, simplified expressions for $\theta_{13}$ can be inferred
\bea
	{\rm NO:}~ h_r=-h_{r,2} & ~\Longrightarrow & \quad \sin^2\theta_{13} \approx \frac{13 + 21\varphi}{3}\,
	\frac{f_r^2}{h_r^2} ~\Longrightarrow \quad f_r \approx \pm h_r/25 \\
	{\rm IO:}~ h_r=+h_{r,2} & ~\Longrightarrow & \quad \sin^2\theta_{13} \approx \frac{3 ( 1 + \varphi)}{5}\, 
	\frac{f_r^2}{h_r^2} ~\Longrightarrow \quad f_r \approx \pm h_r/10.
\eea
In the first case the mass spectrum is dictated by $m_1/m_2 \approx 1$ and $m_3/m_2 \approx 4$ whilst in the second case $m_1/m_2 \approx 1$ and $m_3/m_2 \approx 0$.
The sum of the neutrino masses can be derived using the limits discussed above
\bea
	{\rm NO:} & \quad & \sum_j m_j= \sqrt{\Delta m^2_{31}} \left[ 1.55 - 0.53 
	\sin^2\theta_{13} +\ord(\sin^4\theta_{13}) \right]\approx 7.62 \times 10^{-2}\ \mathrm{eV} \qquad \\
	{\rm IO:} & \quad & \sum_j m_j = \sqrt{-\Delta m^2_{32}} 
	\left[2+10.85\sin^2\theta_{13} +\ord(\sin^4\theta_{13})\right]\approx 1.11 \times 10^{-2}\ \mathrm{eV}, \qquad
\eea
while the effective masses are given by
\bea
	{\rm NO:} & \quad & m_\beta = \sqrt{\Delta m^2_{31}} \left[ 0.26 + 1.73\sin^2\theta_{13} +
	\ord(\sin^4\theta_{13}) \right]\approx 1.47 \times 10^{-2}\ \mathrm{eV} \\
	{\rm IO:} & \quad & m_\beta = \sqrt{-\Delta m^2_{32}} \left[1+ 7.35\sin^2\theta_{13} +\ord(\sin^4\theta_{13})
	\right]\approx 5.75 \times 10^{-2}\ \mathrm{eV}
\eea
and
\bea
	{\rm NO:} & \quad & m_{\beta\beta} = \sqrt{\Delta m^2_{31}} \left[ 0.26 + 0.56\sin^2\theta_{13} +
	\ord(\sin^4\theta_{13}) \right] \approx 1.34 \times 10^{-2}\ \mathrm{eV} \\
	{\rm IO:} & \quad & m_{\beta\beta} = \sqrt{-\Delta m^2_{32}} \left[1+ 6.85 \sin^2\theta_{13} +
	\ord(\sin^4\theta_{13})\right]\approx 5.69 \times 10^{-2}\ \mathrm{eV}.
\eea

%
\subsubsection*{Mechanism II c-2: ${\bf f_r = f_i = 0}$}
\label{sec:typeI_B_xi_small}
%
In this case the Dirac mass is exclusively generated by the flavon in representation $\mbf{5}$.
The mass spectrum, sum rule and mass splittings at LO are the same as in the previous case with $h_i = f_i = 0$ in Section \ref{sec:mechanismII_hi_fi_zero}, so we omit the discussion but recall the conclusions for the ordering: for NO $h_r \approx -h_{r,2}$ and for IO $h_r \approx +h_{r,2}$.
In these limits
\bea
	{\rm NO:}\quad h_r=-h_{r,2} & ~\Longrightarrow & \quad \sin^2\theta_{13} \approx \frac{(7 \varphi +10) }{9} 
	h_i^2/h_r^2 \label{eq:mecIIc_frfi0_NO} \\
	{\rm IO:}\quad h_r=+h_{r,2} & ~\Longrightarrow & \quad \sin^2\theta_{13} \approx \frac{(3 \varphi +2) }{5} 
	h_i^2/h_r^2. \label{eq:mecIIc_frfi0_IO}
\eea
From Eqs.~\eqref{eq:mecIIc_frfi0_NO} and \eqref{eq:mecIIc_frfi0_IO}, we get $h_i/h_r \approx \pm1/10$ for values of $\theta_{13}$ compatible with the experimental determination for both orderings.
The sum of the neutrino masses are
{\small \bea
	{\rm NO:} & \quad & \sum_j m_j=\sqrt{\Delta m^2_{31}} \left[1.55-2.55\sin^2\theta_{13} + 
	\ord(\sin^4\theta_{13})\right] \approx 7.40 \times 10^{-2}\ \mathrm{eV} \quad \\
	{\rm IO:} & \quad & \sum_j m_j=\sqrt{-\Delta m^2_{32}} \left[2+1.58\sin^2\theta_{13} + 
	\ord(\sin^4\theta_{13})\right] \approx 1.01 \times 10^{-1}\ \mathrm{eV}
\eea}
Finally, the effective masses yield
{\small \bea
	{\rm NO:} & \quad & m_\beta = \sqrt{\Delta m^2_{31}} \left[0.26+0.65\sin^2\theta_{13} + 
	\ord(\sin^4\theta_{13})\right]\approx 1.35 \times 10^{-2}\ \mathrm{eV} \\
	{\rm IO:} & \quad & m_\beta = \sqrt{-\Delta m^2_{32}} \left[1+0.65\sin^2\theta_{13} + \ord(\sin^4\theta_{13})
	\right]\approx 5.02 \times 10^{-2}\ \mathrm{eV}
\eea}
and
{\small \bea
	{\rm NO:} & \quad & m_{\beta\beta} =\sqrt{\Delta m^2_{31}} \left[0.26-0.51 \sin^2\theta_{13} + 
	\ord(\sin^4\theta_{13})\right] \approx 1.22 \times 10^{-2}\ \mathrm{eV} \\
	{\rm IO:} & \quad & m_{\beta\beta} =\sqrt{-\Delta m^2_{32}} \left[1+0.15\sin^2\theta_{13} + 
	\ord(\sin^4\theta_{13})\right]\approx 4.96 \times 10^{-2}\ \mathrm{eV}.
\eea}

%
\subsubsection*{Mechanism II c-2: ${\bf h_i = h_r = 0}$}
\label{sec:Z2Z2CP_xr_zero}
%
In this scenario, to get $\theta$ as in Eq.~\eqref{eq:bfp}, we need small $f_{r}$.
Examining the mass spectrum we conclude that this case is only compatible with NO, provided that {\small $h_{r,2} >0 \wedge f_i< -h_{r,2}/2\sqrt{3} $} or {\small $h_{r,2} <0 \wedge f_i> -h_{r,2}/2\sqrt{3}$}.
The sum rule is
\begin{align} \label{sumrule_typeI_C_hr_zero}
	\Sigma &= \left[  \tilde{m}_1 + (21 \varphi +13) \tilde{m}_2 -5 (3 \varphi +2) \tilde{m}_3\right]^2 -(84 
	\varphi +52) \tilde{m}_1 \tilde{m}_2 + \ord(\sin^2\theta_{13}),
\end{align}
which contains a large undisplayed numerical coefficient in front of the $\sin^2\theta_{13}$ term so that large NLO corrections may arise.
For the solar mass splitting we find {\small $\Delta m^2_{21} \propto h_{r,2}(h_{r,2}+2\sqrt{3}f_i)$}, so a natural suppression for $r_1$ is expected when $f_i \approx  -1/2\sqrt{3}h_{r,2}$.
Note that this is also the limit of $z = \ord(f_r^2)$ in the neutrino mass matrix $M_\nu$, as in the previous case. 
Then:
\beq \sin^2  \theta_{13} = \frac{3}{4} (35 \varphi +26) \frac{f_r^2}{h_{r,2}^2} ~\Longrightarrow \quad f_r \approx \pm 1/50\, h_{r,2} \eeq
to reproduce $\sin^2\theta_{13} = 2.206 \times 10^{-2}$.
The mass spectrum is well approximated by $m_1/m_2 \approx 1$ and $m_3/m_2 \approx 9/5$ and additionally we have
\begin{align}
	\sum_j m_j &= \sqrt{\Delta m^2_{31}} \left[2.54-1.97\sin^2\theta_{13}+\ord(\sin^4\theta_{13})\right]\approx 
	1.24\times10^{-1} \ \mathrm{eV} \label{eqn:summjmecIIc2hihr} \nline
	m_\beta &= \sqrt{\Delta m^2_{31}} \left[0.67-0.034\sin^2\theta_{13}+\ord(\sin^4\theta_{13})\right]\approx 
	3.31\times10^{-2} \ \mathrm{eV}
\end{align}
Notice that the value in Eq.~\eqref{eqn:summjmecIIc2hihr} lies in the limit given by \texttt{\large Planck} $\oplus$ BAO in \cite{Aghanim:2018eyx} for the sum of the neutrino masses. 
Finally,
\begin{align} \label{typeI_C_m_betabeta_hr_zero}
	m_{\beta\beta} &= \sqrt{\Delta m^2_{31}} \left[0.67-0.25\sin^2\theta_{13}+\ord(\sin^4\theta_{13})\right]
	\approx 3.29\times10^{-2} \ \mathrm{eV}\,.
\end{align}

%
\subsubsection*{Mechanism II c-2: ${\bf f_r = h_r = 0}$}
\label{sec:case_fr_hr_zero}
%
%
In this case $\theta = \thbf$ in Eq.~\eqref{eq:bfp} implies {\small $h_i \approx \pm h_{r,2} (-0.05 h_{r,2}-0.27 f_i)/(h_{r,2}+0.54  f_i)$}, so small $|h_i|$ is expected.
In this limit, the neutrino mass matrix $M_\nu$ is exactly the same as in the previous case.
At LO the solar and the atmospheric mass differences coincide with those of Section \ref{sec:Z2Z2CP_xr_zero} with $h_i = h_r = 0$.
Also the relation $f_i = k h_{r,2}$, invoked to get the ratio $r_1$ compatible with the data, requires the same $k \approx -1/2\sqrt{3}$.
This in turn implies that
\beq \sin^2\theta_{13} \approx \frac{1}{4} (33 \varphi +25)\, \frac{h_i^2}{h_{r,2}^2} ~\Longrightarrow\quad h_i/h_{r,2} \approx \pm 0.03 \eeq
to get a compatible value of $\theta_{13}$.
Instead, NLO corrections for $m_\beta$ and $m_{\beta\beta}$ are important and make their values different with respect to the case discussed in the case before
\begin{align}
	\sum_j m_j &= \sqrt{\Delta m^2_{31}} \left[2.54-3.94 \sin^2\theta_{13}+\ord(\sin^4\theta_{13})\right]\approx 
	1.22\times10^{-1} \ \mathrm{eV} \nline
	m_\beta &= \sqrt{\Delta m^2_{31}} \left[0.67-0.87\sin^2\theta_{13}+\ord(\sin^4\theta_{13})\right]\approx 
	3.22\times10^{-2} \ \mathrm{eV}
\end{align}
and
\begin{align} \label{typeI_C_m_betabeta_hr_zero}
	m_{\beta\beta} &= \sqrt{\Delta m^2_{31}} \left[0.67-1.10\sin^2\theta_{13}+\ord(\sin^4\theta_{13})\right]\approx 
	3.19\times10^{-2} \ \mathrm{eV}
\end{align}

%
\subsubsection*{Mechanism II c-2: ${\bf h_i = h_{r,2} = 0}$}
\label{sec:Z2Z2CP_xr2_zero}
%
In this realization Eqs.~\eqref{tan2theta_rel} and \eqref{sin13_expression}, expressed in terms of the adimensional parameters in Eq.~\eqref{eqn:mecIIc2AdimParam}, show that $|f_r| \ll 1$ in order to get a small reactor angle.
The LO expressions for the mass differences are related to the ones of the previous case replacing $h_{r,2} \rightarrow h_{r}, f_i \rightarrow - f_i$.
Both orderings are allowed: NO when {\small $h_r >0 \wedge (\varphi -1)/2\sqrt{3} h_r <f_i< h_r/2\sqrt{3} $} or {\small $h_r <0 \wedge (\varphi -1)/2\sqrt{3} h_r>f_i> h_r/2\sqrt{3}$}, and IO for {\small $h_r <0 \;{\rm and}\;  6 f_i > - \sqrt{3}\varphi h_r$} or {\small $h_r >0 \;{\rm and}\;  6 f_i <  -\sqrt{3}\varphi h_r$}.
However, we check in our numerical scan that the IO case is realized at the prize of having all the non-vanishing vevs at the same order of magnitude, which does not have a clear symmetry argument behind it.
The sum rule in this case is
\begin{align} \label{sum_rule_typeI_C_hr2_zero}
	\Sigma &= \left(  \tilde{m}_1 + (34-21 \varphi) \tilde{m}_2 + 5 (3 \varphi -5) \tilde{m}_3\right)^2 + 
	(84 \varphi -136) \tilde{m}_1 \tilde{m}_2 +\ord(\sin^2\theta_{13})
\end{align}
where, as in the case $h_i=h_r=0$, the coefficient in front of $\sin^2\theta_{13}$ is large and we have to contemplate significant NLO corrections.
Also to obtain a relation among $h_{r}$ and $f_i$, NLO terms to the ratio $r_1$ expressed as a series in $\sin^2\theta_{13}$ are needed to be considered.
Fixing both $r_1$ and $\theta_{13}$ to their best fit values we get $h_r = 2\sqrt{3} f_i$.
In this limit, the NO spectrum is dictated by $m_1/m_2 \approx 1$ and $m_3/m_2 \approx 9/5$ and the reactor angle is approximated by 
\beq \sin^2 \theta_{13} \approx \frac{3 (119 \varphi +74)}{4}\,\frac{f_r^2}{h_r^2} ~\Longrightarrow\quad f_r \approx \pm h_r/100  \label{eqn:mecIIc2hihr2cond}\eeq
to obtain a value compatible with the data. Finally, still considering the relation in Eq.~\eqref{eqn:mecIIc2hihr2cond} for the vevs, we get
\begin{align}
	\sum_j m_j &\approx \sqrt{\Delta m^2_{31}} \left[2.54-3.03 \sin^2\theta_{13} +\ord(\sin^4\theta_{13})\right] 
	\approx 1.23\times10^{-1} \ \mathrm{eV} \label{eqn:summecIIc2hihr2} \nline
	m_\beta & \approx \sqrt{\Delta m^2_{31}} \left[0.67-0.50\sin^2\theta_{13}+\ord(\sin^4\theta_{13})\right]
	\approx 3.26\times10^{-2} \ \mathrm{eV} \nline
	m_{\beta\beta} & \approx \sqrt{\Delta m^2_{31}} \left[0.67-0.71\sin^2\theta_{13}+\ord(\sin^4\theta_{13})
	\right]\approx 3.23\times10^{-2} \ \mathrm{eV} \label{typeI_C_m_betabeta_hr2_zero}\,.
\end{align}
Note that the prediction for the sum of the neutrino masses in Eq.~\eqref{eqn:summecIIc2hihr2} is in the limit provided by \texttt{\large Planck} $\oplus$ BAO in \cite{Aghanim:2018eyx}.

%
\subsubsection*{Mechanism II c-2: ${\bf f_r = h_{r,2} = 0}$}
\label{sec:case_fr_hr2_zero}
%
In this case $M_\nu$ has the same LO structure as the previous case in Section \ref{sec:Z2Z2CP_xr2_zero} with $h_i=h_{r,2}=0$.
Equal predictions for the mass spectrum and $\Sigma$ are obtained.
As before, a natural suppression of the ratio $r_1$ is achieved for $f_i = k h_r$ with $k \approx 1/2\sqrt{3}$.
Under this ansatz, the reactor mixing angle is
\beq 
	\sin^2\theta_{13} \approx \frac{13-3 \varphi }{4} h_i^2/h_r^2 ~\Longleftrightarrow\quad h_i \approx \pm h_r/10
\eeq
to obtain $\sin^2\theta_{13} = 2.206 \times 10^{-2}$.
In this specific limit, the mass spectrum is constrained to be $m_1/m_2 \approx 1$ and $m_3/m_2 \approx 9/5$.
The value of the total sum of the neutrino masses and the effective masses are
\begin{align}
	\sum_j m_j &= \sqrt{\Delta m^2_{31}}\left[2.54-7.64 \sin^2\theta_{13} +\ord(\sin^4\theta_{13})\right]\approx 
	1.18 \times 10^{-1}\ \mathrm{eV} \nline
	m_\beta &= \sqrt{\Delta m^2_{31}}\left[0.67-2.28  \sin^2\theta_{13} +\ord(\sin^4\theta_{13})\right]\approx 3.06 
	\times 10^{-2}\ \mathrm{eV}
\end{align}
and
\begin{align} \label{typeI_D_mbetabeta_hr2_zero}
	m_{\beta\beta} &=\sqrt{\Delta m^2_{31}}\left[0.67-2.50  \sin^2\theta_{13} +\ord(\sin^4\theta_{13})\right]\approx 
	3.04 \times 10^{-2}\ \mathrm{eV}\,.
\end{align} 

\section{Numerical results}
\label{sec:numerics}
%
In this section we study numerically the realizations analysed before.
We inspect the accuracy of the results presented in Section \ref{sec:analytics}, which have been computed using perturbation theory, and provide additional information about the allowed range of values for the effective masses.\\
The numerical scan is performed as follows: 
first, the parameters of $M_\nu$ are randomly generated with a flat distribution in the range $[-1, +1]$; 
then, for those points whose observables remain in the $3\sigma$ region, we compute the $\chisquare$ for the permitted orderings, as defined in Appendix \ref{sec:chisquare}, with the additional constraint {\small $\sum_j m_j \leq 0.12\ \mathrm{eV}$}\footnote{As indicated throughout the text, some of the studied cases present some incompatibility with the constraint from \texttt{Planck} $\oplus$ BAO over the sum of the light neutrino masses.
For completeness, we keep their results in the tables and comment it when opportune.}; 
finally, we extract the minimum for each case and collect all the results in Tables \ref{tab:mechanismI_chisquare}$-$\ref{tab:mechanismIIc2_chisquare}.
We also report the $\chisquare$ for those cases which could not be expressed as a series in the natural small parameter and we marked them with $\xmark$.
The contributions due to the mixing angles $\chisquare_a$ and the mass splittings $\chisquare_m$ are specified too.

\subsection{Accuracy of the perturbative results}
\label{subsec:accuracy}
%
The goodness of the perturbative expansions are evaluated using the following {\it associated error} defined for a given observable $q$:
\begin{align}
	\Delta q \equiv \frac{q^{\mathrm{Full}}- q^{\mathrm{(N)LO}}}{q^{\mathrm{Full}}}.
\end{align}
The superscript Full refers to the full numerical evaluation obtained in the numerical scan while ${\rm (N)LO}$ refers to the order of the analytical expansion of the observable $q$.
We comment the obtained results for each case.

\subsubsection*{Mechanism I}
The correlations between the parameters in the cases $z=0,~x=0$ and $s=0$ are confirmed to a very good accuracy.
The perturbative estimates of the reactor angle and the ratio $r_\ell$ in all three cases show a good agreement with the full numerical results at the level of $\Delta \approx 10\%- 20\%$.

\subsubsection*{Mechanism II a-1}
For the case $Z = 0$ we confirm the existence of the two correlations $S \approx X$ and $|S| \gg |X|$. The corrections to the analytical expression of $r_1$ and $\theta_{13}$ given in Section \ref{sec:typeI_Z_zero} turns out to be roughly 10\% for both observables. For the case $X = 0$,  the three different correlations between $S$ and $Z$ have also been found numerically. We do not observe  any specific distribution of good points in the plane $(r_1, \sin^2\theta_{13})$. This also reflects on the fact that the NLO corrections, especially to $r_1$, are not completely negligible; in fact, $|\Delta r_1| \approx 50\%$.

\subsubsection*{Mechanism II a-2}
For all subcases discussed in the analytic part, there is a good agreement with the estimate of $\sin^2\theta_{13}$ at the level of  $\approx 10\%$.
There is also a general trend of a not so good perturbative estimate for both $\Delta m^2_{21}$ and $r$; the comparison with the numerical computation shows that the LO estimates are generally off by more that 50\% and only the inclusion of NLO terms (always giving rise to cumbersome expressions, not shown in this paper) reduces these discrepancies down to $\approx$ 10\% - 20\%.
Finally, the numerical simulation reveals that {\small $h_i=h_{r,2}=0$} with NO is also possible provided that $|h_r|\ll |g|\ll |f|$.
However, this hierarchy is not justified by any symmetry argument.

\subsubsection*{Mechanism II c}
As a general remark, the LO expressions found for $\theta_{13}$ are in very good agreement with the numerical results,  the largest discrepancy being $\Delta \sin^2\theta_{13}\approx 30-45 \%$ for the case $f_r = h_{r,2} = 0$ whereas $r_\ell$ suffers from large corrections in all studied cases; the discrepancy at LO is at the 60\% level on average (but only $\approx 5-15 \%$ for the case $f_r = h_{r,2} = 0$) and it is strongly reduced after NLO corrections are taken into account, where only a small 5-10 \% discrepancy remains.

\subsection{Predictions for \mathversion{bold} $m_\beta$ and $m_{\beta\beta}$ \mathversion{normal}}
\label{subsec:mbmbb}
%
In this section we report the numerical estimates for $m_\beta$ and $m_{\beta\beta}$ for each of the mechanisms studied before. 
We show $\mbetabeta$ as a function of $m_{\min}$ and $\mbeta$. 
The excluded region for $\mbetabeta$ in both planes  corresponds to $\mbetabeta \geq 0.11\ \mathrm{eV}$ @$90\%$ CL, a limit obtained from the decay half-time of $^{130}$Te with the combined results of \texttt{\large CUORE} \cite{Alduino:2017ehq}, \texttt{\large CUORE-0} \cite{Alduino:2016zrl} and \texttt{\large Cuoricino} \cite{Andreotti:2010vj} experiments.
A similar limit can be read from the \texttt{\large GERDA} experiment \cite{Agostini:2018tnm}: $\mbetabeta \geq 0.12\ \mathrm{eV}$ @$90\%$ CL.
Future data from \texttt{\large CUORE} and \texttt{\large SNO+} ($^{130}\mathrm{Te}$), \texttt{\large GERDA} and \texttt{\large MAJORANA} ($^{76}\mathrm{Ge}$), \texttt{\large SuperNEMO}, \texttt{\large KamLAND-Zen} and \texttt{\large EXO} ($^{136}\mathrm{Xe}$), \texttt{\large AMoRE} and \texttt{\large MOON} ($^{100}\mathrm{Mo}$), \texttt{\large COBRA} ($^{116}\mathrm{Cd}$), \texttt{\large CANDLES} ($^{48}\mathrm{Ca}$) could probe the IO and quasi-degenerate region with a sensitivity $\mbetabeta \approx (0.01 \div 0.05)\ \mathrm{eV}$ (for reviews see \cite{deSalas:2018bym} and \cite{Dell'Oro:2016dbc}).
The bounds on $m_{\min}$ can be obtained from the recent analysis of the \texttt{\large Planck} collaboration \cite{Aghanim:2018eyx}: $m_{\min} < 0.09\ \mathrm{eV}$ (\texttt{\large Planck}) and $m_{\min} < 0.04 \ \mathrm{eV}$ (\texttt{\large Planck} $\oplus$ BAO).
For $\mbeta$, we indicate with a vertical red dashed line the expected sensitivity for the \texttt{\large KATRIN} experiment: $0.2\ \mathrm{eV}\ @\ 90\% \ \mathrm{CL}$ \cite{Osipowicz:2001sq}.

\subsubsection*{Mechanism I} \label{sec:mbetabeta_mechanism_I}
\begin{figure}[h!]
	\vspace{-0.4cm}
	\centering
	\includegraphics[scale=.412]{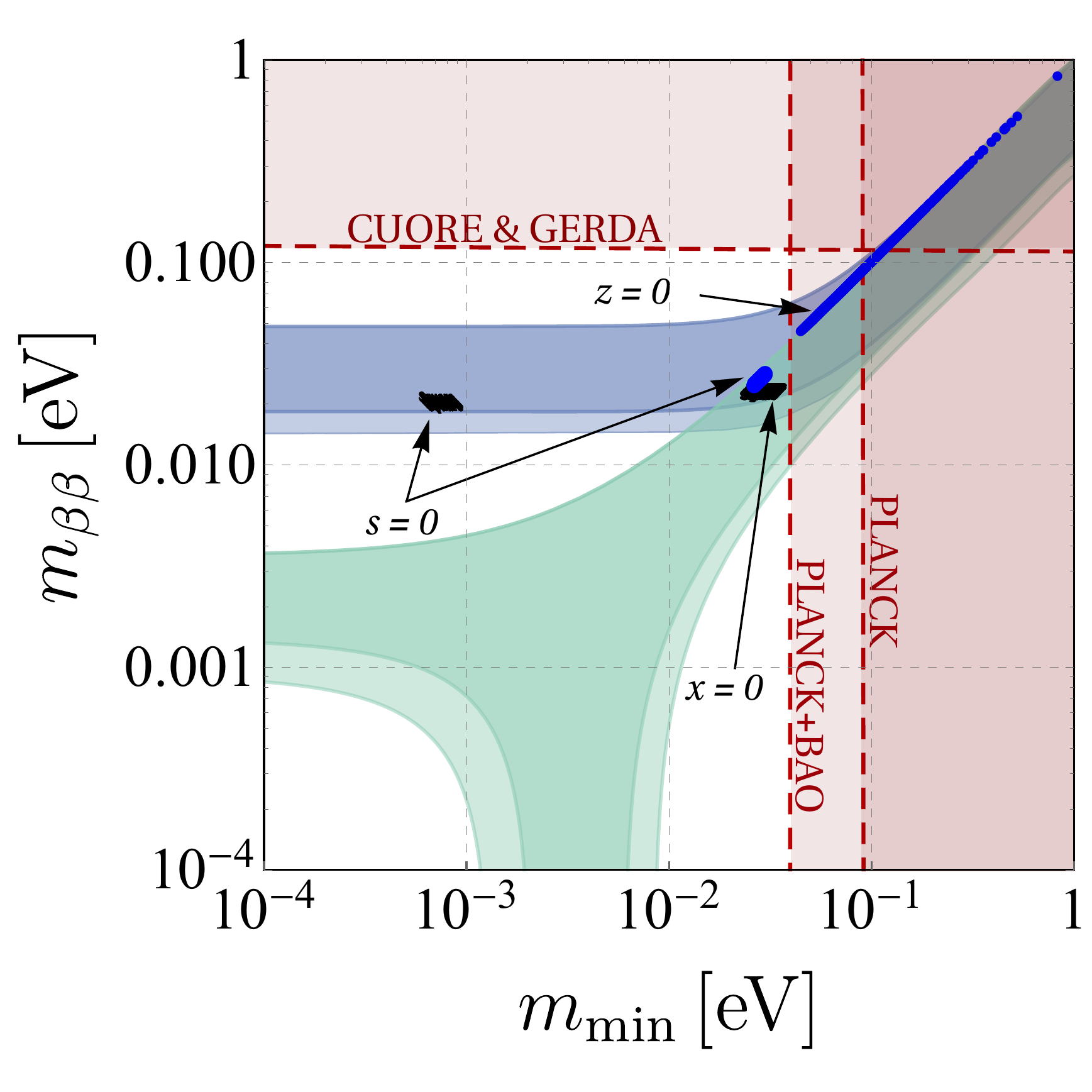}
	\includegraphics[scale=.418]{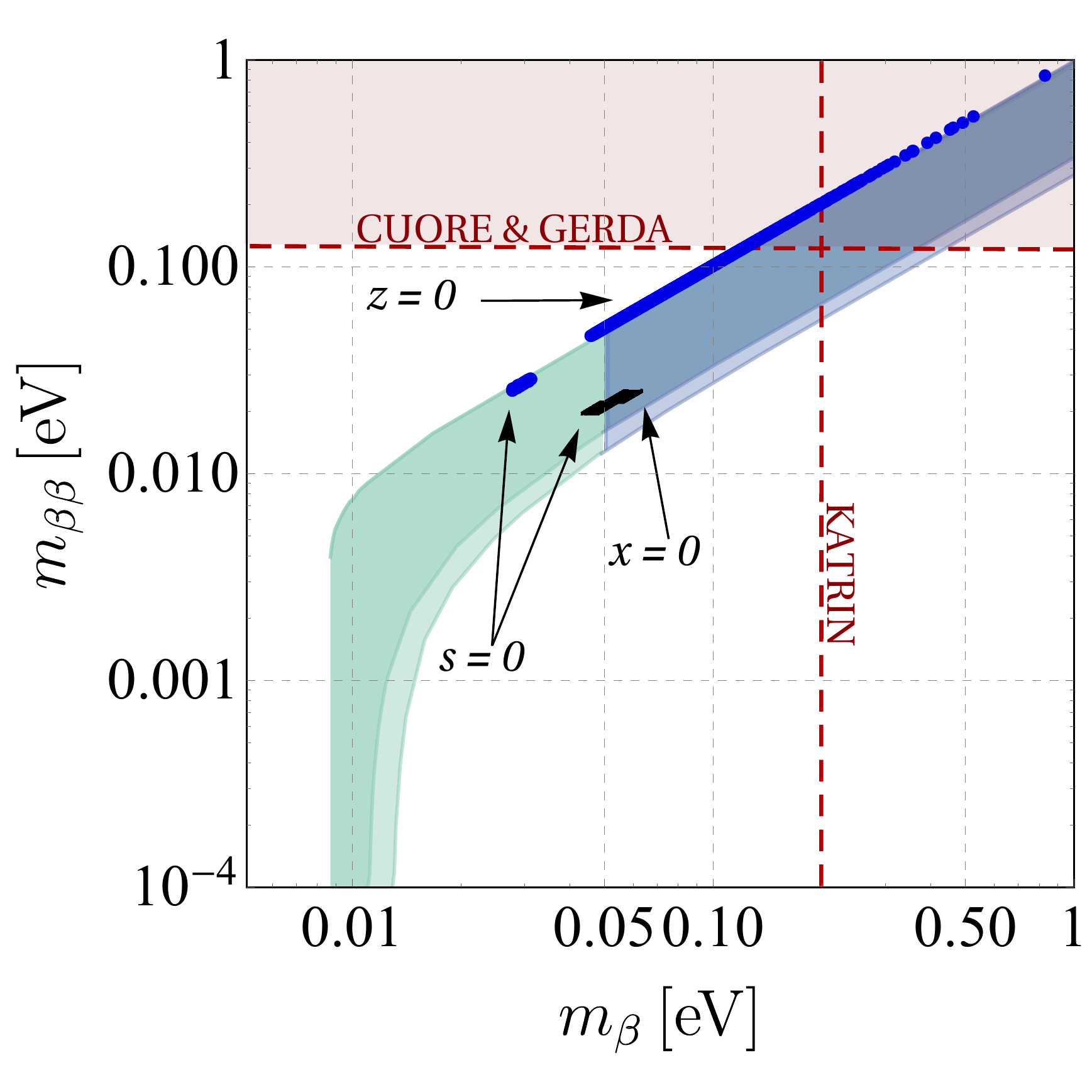}
	\caption{\label{fig:weinberg_m_betabeta} \it \small
	Effective mass $\mbetabeta$ for the neutrinoless double beta decay as function of $m_{\mathrm{min}}$ (left 
	plot) and $m_{\beta}$ (right plot) in the case of the Weinberg operator for $z =0$, $x= 0$ or $s = 0$.
	Blue circles are for NO while black diamonds for IO.
	The green (blue) region is the allowed area for $m_{\beta \beta}$ at $3\sigma$ CL of mixing parameters assuming NO (IO) while the green (blue) lines contain the region at $1\sigma$.}
\end{figure}

Figure \ref{fig:weinberg_m_betabeta} shows $\mbetabeta$ as a function of $m_{\min}$ (left panel) and $\mbeta$ (right panel) assuming $s= 0$, $x = 0$ or $z = 0$.
Our analytic results in Section \ref{subsec:mechanism_I} are in agreement with the numerical evaluation.
We also show the results for the case $s = 0$ assuming NO, even though a natural perturbative expansion in the small parameter $y$ was not possible.
We confirm that the predictions in case $z = 0$ are not compatible with current data on neutrinoless double decay (high mass region) \cite{Alduino:2017ehq} and the latest combined limit from \texttt{\large Planck} $\oplus$ BAO (low mass region) \cite{Aghanim:2018eyx}.
Also the case $x=0$ is not consistent with the \texttt{\large Planck} $\oplus$ BAO limit on the sum of the light neutrino masses.
On the other hand, $s=0$ (NO) could be proved in the near future with more stringent bounds coming from cosmology.
Similarly, future sensitivity of $\mbetabeta \approx (0.01 \div 0.05)\ \mathrm{eV}$ will allow us to confirm or reject the $s=0$ (IO) realization.
Conversely, our preditions for $\mbeta$ are far from the expected reach of the \texttt{\large KATRIN} experiment.\\

\begin{figure}[h!]
	\centering
	\includegraphics[scale=.415]{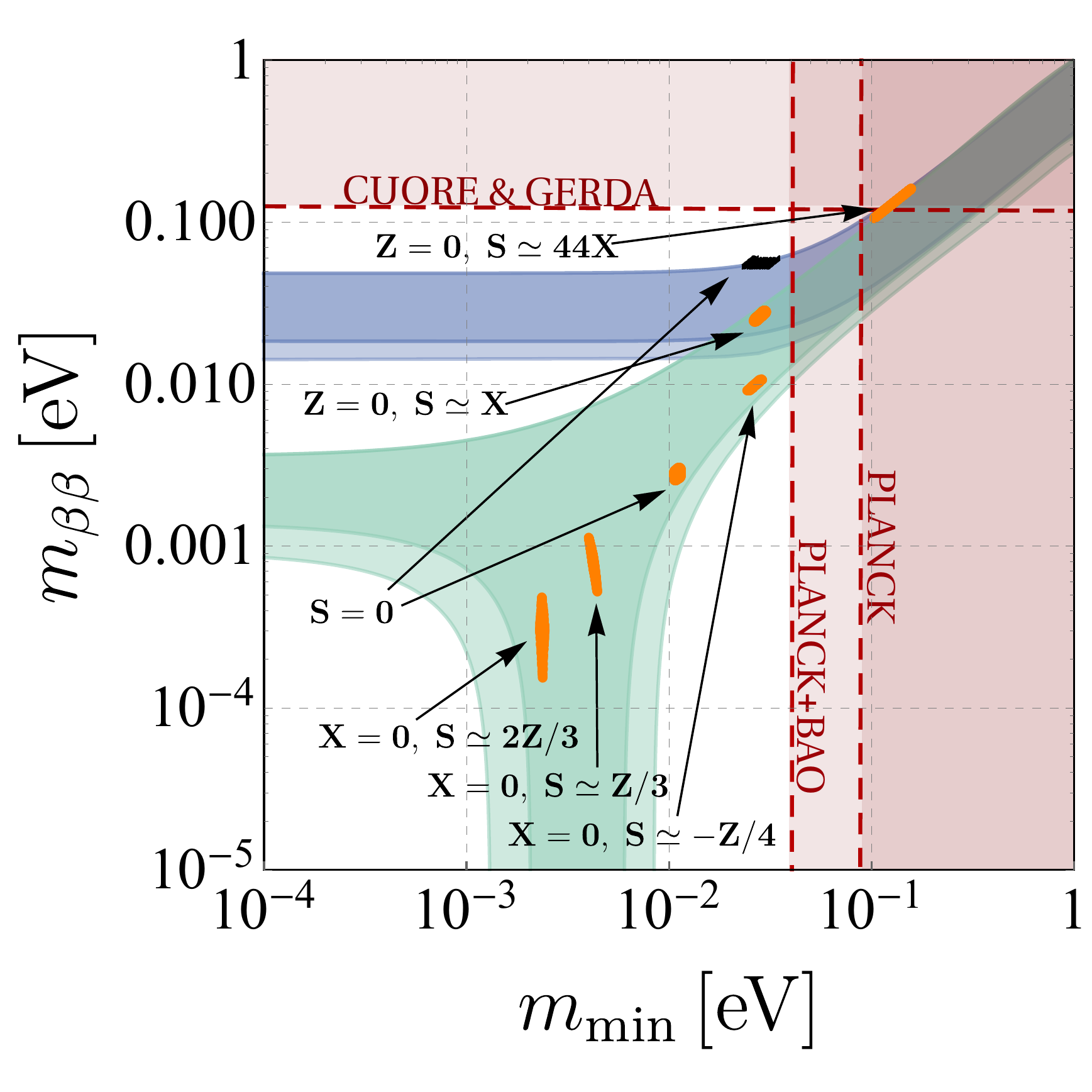}
	\includegraphics[scale=.425]{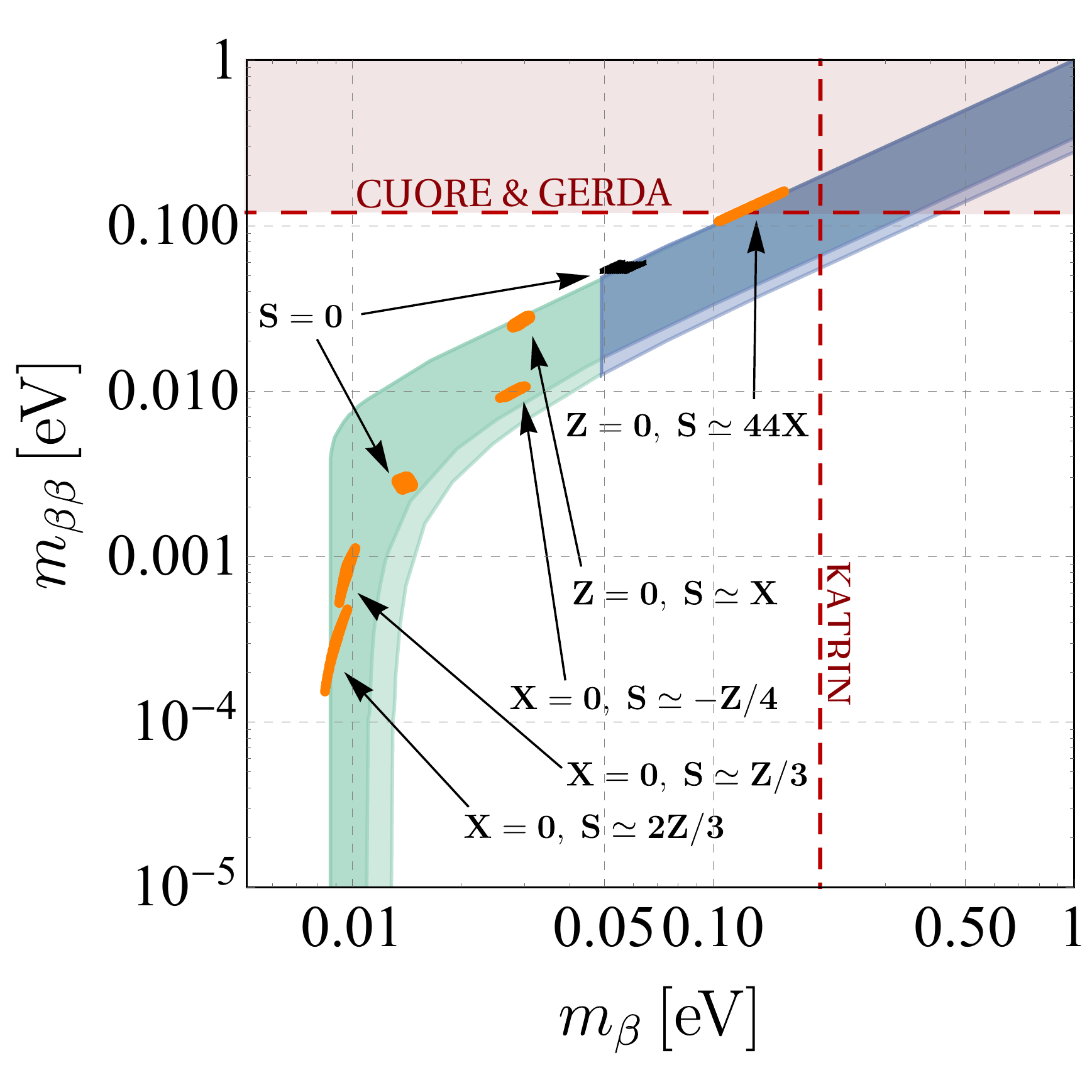}
	\caption{ \label{fig:mechanismII_a1_m_betabeta} \it \small
	The same as Figure \ref{fig:weinberg_m_betabeta} in the case of Mechanism II a-1: $Z = 0$ with $S\approx 	
	\lbrace X,\,44X\rbrace$, $S = 0$ and $X = 0$ with $S \approx \lbrace -Z/4,\, Z/3,\, 2Z/3\rbrace$.
	Orange circles are for NO while black diamonds for IO.}
\end{figure}
\newpage
\begin{figure}[h!]
	\centering
	\includegraphics[scale=.425]{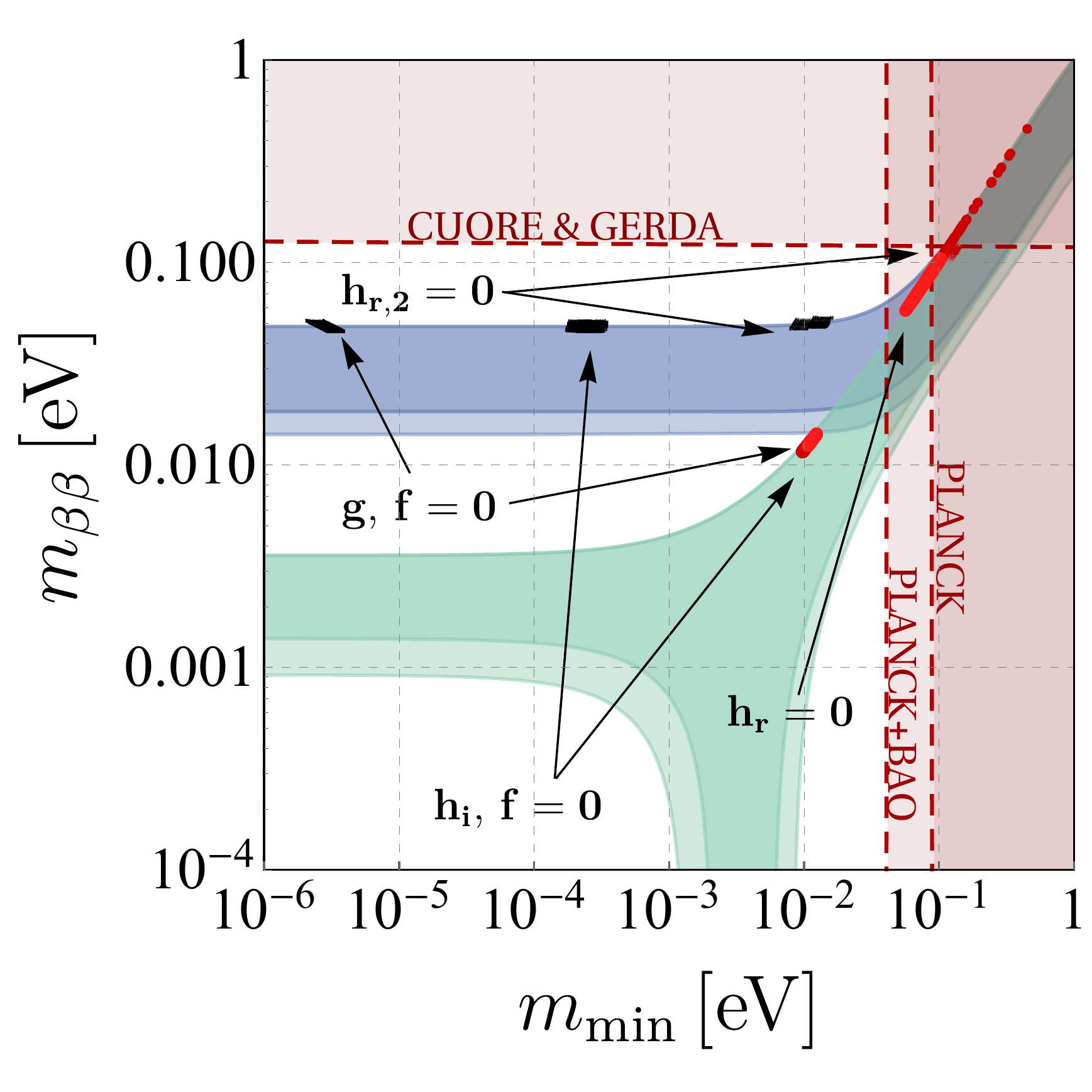}
	\includegraphics[scale=.425]{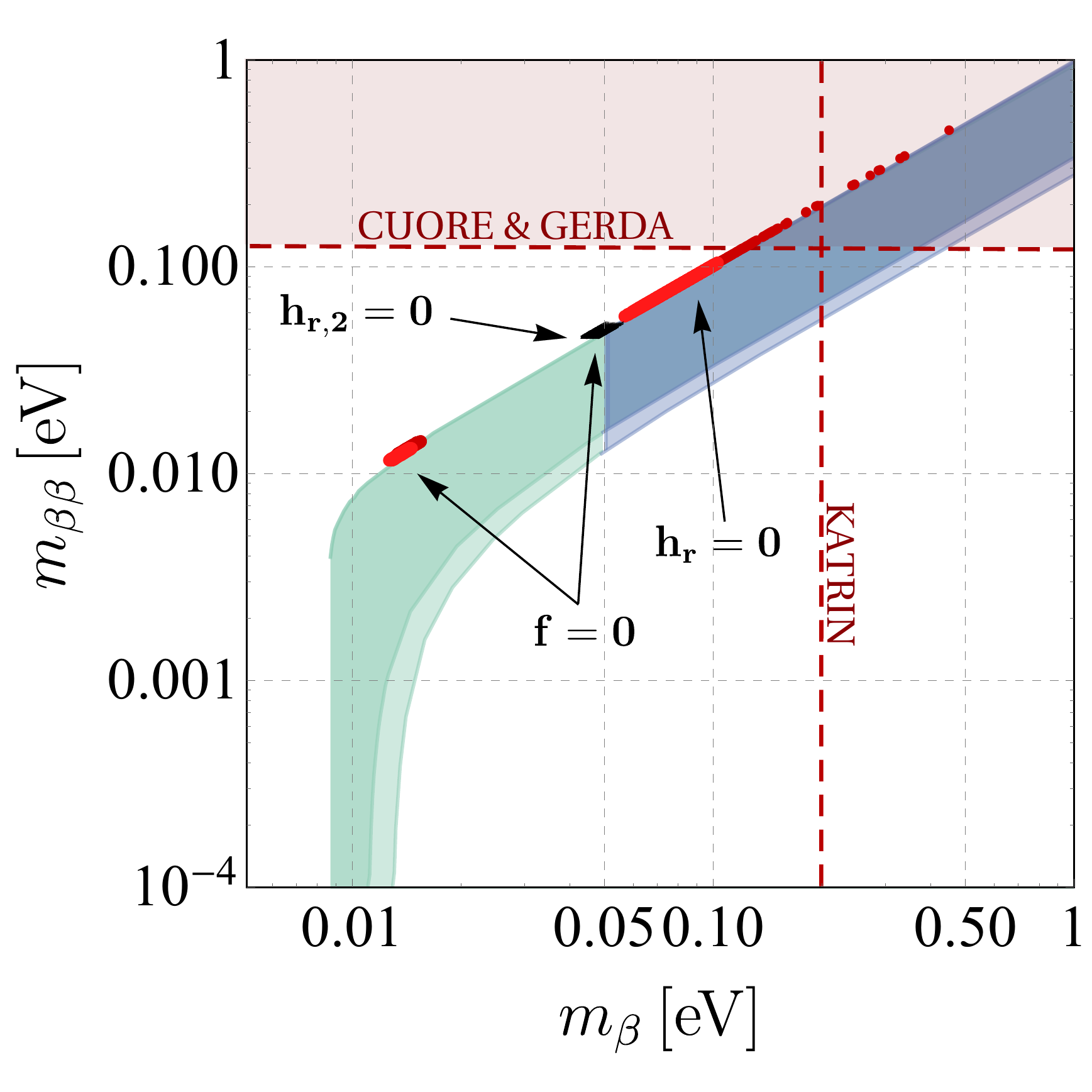}
	\caption{ \it \small \label{fig:mechanism_II_a2_m_betabeta}
	The same as Figure \ref{fig:weinberg_m_betabeta} in the case of Mechanism II a-2.
	Red dark (light) circles correspond to NO and $h_i=0$ ($g=0$) while black diamonds to IO.}
\end{figure}
\begin{figure}[h!]
	\centering
	\includegraphics[scale=.425]{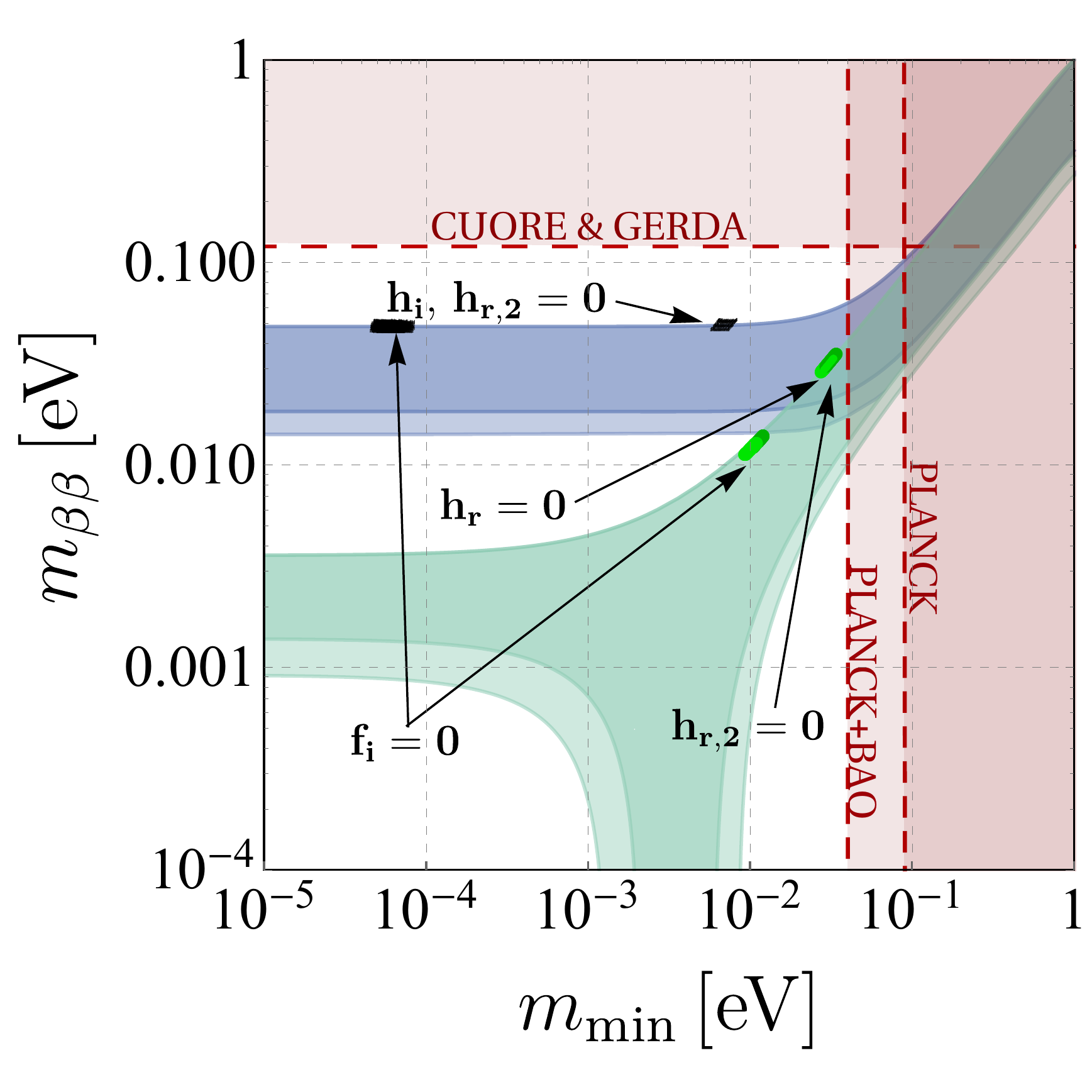}
	\includegraphics[scale=.425]{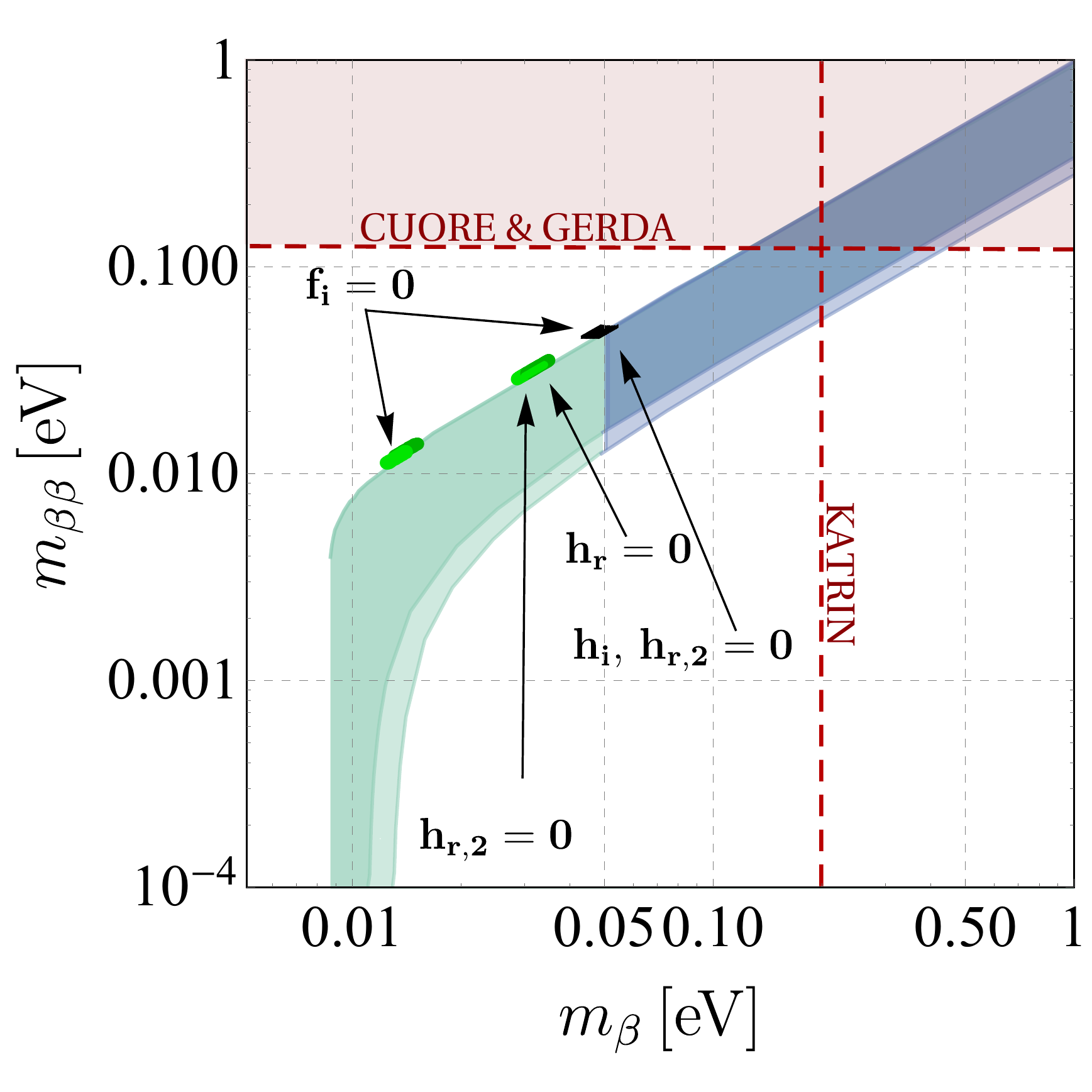}
	\caption{ \it \small \label{fig:mechanismII_c_m_betabeta}
	The same as Figure \ref{fig:weinberg_m_betabeta} in the case of Mechanism II c-2.
	Green dark (light) circles are for NO with $h_i=0$ ($f_r=0$) while black diamonds for IO.}

\end{figure}

\subsubsection*{Mechanism II a-1}
Figure \ref{fig:mechanismII_a1_m_betabeta} is devoted to Mechanism II a-1, where  we also report the case {\small $S = 0$} that was not discussed analytically since it is not related to patterns with a natural hierarchy among the vevs.
The predictions for the various cases are in agreement with our numerical scan.
The case {\small $Z = 0$} with {\small $|S| \gg |X|$}, part of which lies in the $m_{\beta\beta}$ exclusion region, can be completely discarded making use of the latest \texttt{\large Planck} data \cite{Aghanim:2018eyx}, as already outlined in Section \ref{sec:typeI_Z_zero}.
Similarly, we find out that {\small $S = 0$} (IO) is not compatible with the combined bound from \texttt{\large Planck} $\oplus$ BAO.
Instead, the results obtained for {\small $Z = 0$} with {\small $S\simeq X$, $X=0$} and {\small $S\simeq -Z/4$} could be probed in the following years by neutrinoless double beta decay experiments and improved data from cosmology.
The rest of the cases associated with {\small $X = 0$} and {\small $S=0$} (NO) are beyond the expected sensitivity.
For $\mbeta$, only the case {\small $|S| \gg |X|$} is close to the \texttt{\large KATRIN} sensitivity.

\subsubsection*{Mechanism II a-2} \label{sec:mbb_mechanismII_a2}
Mechanism II a-2 contains six different realizations, although the cases with $g=0$ or $h_i=0$ are almost degenerated in their predictions for the effective masses when considering $h_r=0$ and $h_{r,2}=0$.
Then, we omit them in the labelling.
Indeed, experimentally, these two pairs of situations might not be distinguishable.
We observe that the case $h_r=0$ gives rise to results that are similar to those from $z=0$ in Mechanism I.
Current limits from \texttt{\large Planck} $\oplus$ BAO are not compatible with this scenario.
On the other hand, future sensitivity in $\mbetabeta$ could help in proving the cases that predict IO, namely those involving $f=0$ and $h_{r,2}=0$.
All the analytical predictions discussed in Section \ref{subsec:mechanism_IIa2} are very well confirmed 
by these numerical results.

\subsubsection*{Mechanism II c}
This mechanism, discussed in detail in Section \ref{subsec:mechanism_IIc}, contains six possible set of the vacuum alignments, although considering $h_i=0$ or $f_r=0$ give almost equal predictions for the effective masses.
Therefore we omit them in the labelling.
The results  (in agreement with our analytical predictions) are shown in Figure \ref{fig:mechanismII_c_m_betabeta}.
We observe that further data from cosmology and $0\nu\beta\beta$ decay experiments could be sensitive enough to prove or discard those realizations for $h_r=0$ and $h_{r,2}=0$, which predicts NO.
Also the scenario with $f_i=0$ (IO) could be tested with future data from neutrinoless double beta decay.
On the other hand, our estimations for $m_\beta$ remain far from the expected sensitivity of \texttt{\large KATRIN}.

%
\section{Summary}
\label{sec:summary}
%
We have analysed all possible neutrino mass realizations under the flavour group $A_5  \times {\rm CP}$ spontaneously broken to two residual symmetries in the neutrino, ${\cal G}_\nu=Z_2\times {\rm CP}$, and charged-lepton sector, ${\cal G}_e=Z_5$. This framework reproduces the phenomenology of Case II in \cite{DiIura:2015kfa}.
In the considered basis, the charged leptons are diagonal and the $U_{\rm PMNS}$ matrix is exclusively given by the rotation to the mass basis of the neutrino states.
Then, the structure of neutrino mass matrix, the $U_{\rm PMNS}$ and the flavon vevs are completely fixed by the residual symmetry and have been computed according to it in Sections \ref{subsec:review} and \ref{subsec:flavon_vevs}.\\
In Section \ref{sec:analytics} we have derived analytic estimates for the most important parameters related to neutrino phenomenology: mass spectrum, squared mass differences, sum rule, mass ratio, effective masses and Majorana phases. When possible, simplified relations among parameters computed with perturbation theory were provided, in order to capture the dominant effects. The allowed orderings and correspondent relations among vevs have also been presented. This has been performed for a set of realizations that cover all possible ways of generating neutrino masses, under the assumption of single type of particles added to the SM. They are: Mechanism I, which groups the Weinberg operator and type II see-saw, and Mechanism II, consisting of type I and III see-saw. The last case was divided in two subcases: Mechanism II-1, considering a trivial Dirac mass matrix, and Mechanism II-2, where the Majorana mass matrix was trivial. Each of these cases was analysed putting one or more flavon vevs equal zero. This approach is equivalent to leave out some of the flavons in a model or, if not, it can be always arranged through a suitable vacuum alignment. \\
Finally, Section \ref{sec:numerics} was devoted to a numerical scan where the {\small $\chi^2$} analysis was performed. The results have been shown in Tables \ref{tab:mechanismI_chisquare}$-$\ref{tab:mechanismIIc2_chisquare} and the contributions coming from the mixing angles and mass splittings have been specified. The observed common trend was a significantly worst fit for the mixing angles compared to the one for the mass differences. In Section \ref{subsec:accuracy}, the accuracy of the analytical expressions derived before has been evaluated and we have found a general good agreement, except for some cases where the inclusion of NLO corrections was essential to reduce the discrepancies with the numerical predictions. Finally, Section \ref{subsec:mbmbb} has been devoted to display our estimations for the effective masses $m_\beta$ and $m_{\beta\beta}$. We notice that some realizations are expected to be tested thanks to $0\nu\beta\beta$ experiments and cosmological constraints in the next years. On the other hand, as already expected from the analytic results, Mechanism II a-2 and II c-2 showed a great degeneracy in the predictions of some of the cases and the prospect of distinguishing among them relies on having an improved knowledge of $m_{\rm min}$ from future cosmological experiments.

%
\section*{Acknowledgments}
%
We thank Claudia Hagedorn for the help in the early stages of this work and for checking some of the analytical results presented in this paper.
AD acknowledges the hospitality of the Excellence Cluster Universe where part of this work was carried out.\\

{\small {\bf Note Added}: At the end of the completion of this work a new global fit from the NuFIT collaboration has been released \cite{Esteban:2018azc}.
Our results for the $\chi^2$ analysis are consistent with the new data, barring a slightly worse goodness of fit due to the new preferred value of $\theta_{23}$.
The symmetry analysed here predicts a maximal atmospheric angle and therefore, with the new results, our results in Tables \ref{tab:mechanismI_chisquare}-\ref{tab:mechanismIIc2_chisquare} receive an additional contribution $\chi^2_{\sin^2\theta_{23}}\sim 15$.
However, we do not expect this to be a major problem when constructing specific models, for which NLO corrections should be considered and may perfectly account for this difference.
We have also checked that our estimations remain approximately the same when considering other global fits such as those presented in \cite{deSalas:2017kay} and \cite{Capozzi:2018ubv}.}

\newpage
%
\appendix
\section{Group Theory of ${\bf A_5\times}$ {\bf CP}}
\subsection{Generators} \label{app:generators}
Here we show the generators of the $A_5$ group, $s$ and $t$, in the ${\bf 1}$, ${\bf 3'}$, ${\bf 4}$ and ${\bf 5}$ representations.
	\begin{subequations} \label{eqn:genrep}
	\beq \label{eqn:gen1rep} S_{\bf 1} \;=\; e^{i\, \pi} \hspace{1.cm} T_{\bf 1} \;=\; e^{i\, 
		\frac{2\pi}{5}} \eeq
		
	\beq \label{eqn:gen3prep} S_{\bf 3'} = -\frac{1}{\sqrt{5}}\begin{pmatrix}
		    1      & \sqrt{2}    & \sqrt{2}   \\
		~\sqrt{2}~ & ~1/\varphi~ & ~-\varphi~ \\
		  \sqrt{2} & -\varphi    & 1/\varphi  \\		
		\end{pmatrix}
		\hspace{1.cm} T_{\bf 3'} = \begin{pmatrix}
	 	    1 &         0                & 0                         \\
		  ~0~ & ~e^{i\, \frac{4\pi}{5}}~ & 0                         \\
		    0 &         0                & ~e^{-i\, \frac{4\pi}{5}}~ \\		
		\end{pmatrix}  \eeq
		
	\beq \label{eqn:gen4rep} S_{\bf 4} = -\frac{1}{5}\begin{pmatrix}
	 	~-\sqrt{5}~ & ~\varphi-3~ & ~\varphi+2~ & ~-\sqrt{5}~ \\
		  \varphi-3 &    \sqrt{5} &    \sqrt{5} &   \varphi+2 \\
		  \varphi+2 &    \sqrt{5} &    \sqrt{5} &   \varphi-3 \\		
		  -\sqrt{5} &   \varphi+2 &   \varphi-3 &   -\sqrt{5} \\
		\end{pmatrix}
		\hspace{0.1cm} T_{\bf 4} = \begin{pmatrix}
		~e^{i\, \frac{2\pi}{5}}~ &         0                &          0                &            0            \\
		           0             & ~e^{i\, \frac{4\pi}{5}}~ &          0                &            0            \\		
		           0             &         0                & ~e^{i\, \frac{6\pi}{5}}~  &            0            \\
		           0             &         0                &          0                & ~e^{i\, \frac{8\pi}{5}}~\\		
		\end{pmatrix} \eeq
		
	\beq \label{eqn:gen5rep} \hspace{-0.35cm} S_{\bf 5} = {\scriptsize \frac{1}{5}\begin{pmatrix}
	 	           ~~~-1 &  \sqrt{6}      & -\sqrt{6}     &  -\sqrt{6}    & -\sqrt{6}    \\
	 	     ~~~\sqrt{6} & ~2-\varphi~    & 2\varphi      & 2(1-\varphi)~ & -(1+\varphi) \\	 	
		     ~-\sqrt{6}~ &  2\varphi      & 1+\varphi     &   2-\varphi   & 2\,(\varphi-1) \\
		       -\sqrt{6} & 2\,(1-\varphi) & 2-\varphi     & 1+\varphi     & -2\varphi    \\		
		       -\sqrt{6} & -(1+\varphi)   & ~2(\varphi-1) & -2\varphi     & 2-\varphi    \\
		\end{pmatrix}}
		\hspace{0.25cm} T_{\bf 5} = {\scriptsize \begin{pmatrix}
		~1~ &            0             &         0                &          0                &            0            \\
		 0  & ~e^{i\, \frac{2\pi}{5}}~ &         0                &          0                &            0            \\
		 0  &            0             & ~e^{i\, \frac{4\pi}{5}}~ &          0                &            0            \\		
		 0  &            0             &         0                & ~e^{i\, \frac{6\pi}{5}}~  &            0            \\
		 0  &            0             &         0                &          0                & ~e^{i\, \frac{8\pi}{5}}~\\		
		\end{pmatrix}} \eeq
	\end{subequations}
	
For $X_{0,{\bf r}}$ in the representation ${\bf r}\in A_5$ we consider:
	\begin{subequations} \label{eqn:X0rep}
	\beq X_{0,{\bf 1}} \;=\; {\bf 1} \hspace{2.cm}
		 X_{0,{\bf 3'}} = \begin{pmatrix}
	 	    								~1~ & ~0~ & ~0~ \\
	 	    							 	 0  &  0  &  1  \\
	 	    								 0  &  1  &  0
	 	    								\end{pmatrix}
	\eeq		
	\beq X_{0,{\bf 4}} = \begin{pmatrix}
	 					~0~ & ~0~ & ~0~ & ~1~ \\
	 					 0  & ~0~ & ~1~ & ~0~ \\
	 					 0  & ~1~ & ~0~ & ~0~ \\
	 					 1  & ~0~ & ~0~ & ~0~
	 					\end{pmatrix} \hspace{1.cm}
	 	 X_{0,{\bf 5}} = \begin{pmatrix}
	 					~0~ & ~0~ & ~0~ & ~0~ & ~1~ \\
	 					~0~ &  0  & ~0~ & ~1~ & ~0~ \\
	 					~0~ &  0  & ~1~ & ~0~ & ~0~ \\
	 					~0~ &  1  & ~0~ & ~0~ & ~0~ \\
	 					~1~ &  0  & ~0~ & ~0~ & ~0~
	 					\end{pmatrix}
	\eeq
	\end{subequations}	

\subsection{Kronecker products of $A_5$} \label{sec:KP_A5}
%
We report here the complete list of the Kronecker products for the group $A_5 \times CP$. With respect to the simple group $A_5$ we need the additional condition
\begin{align}
 \bigg[X(\mbf{r} \otimes \mbf{r'})^*\bigg]_{\mbf{r''}} = \bigg[X(\mbf{r})^*\otimes X(\mbf{r'})^*\bigg]_{\mbf{r''}} \qquad \forall \ \mbf{r}, \mbf{r'}, \mbf{r''} \in A_5
\end{align}
where $X$ is the $CP$ matrix for the representation $\mbf{r} \in A_5$.
In the following we report only the Kronecker products in the case of $X = X_0$ which differ from the results quoted in \cite{Feruglio:2011qq}.
We assign $a=(a_1,a_2,a_3)^T$ and $b=(b_1,b_2,b_3)^T$ to the $\mbf{3}$ representation, while $a'=(a_1',a_2',a_3')^T$ and $b'=(b_1',b_2',b_3')^T$ belong to the $\mbf{3'}$ representation, $c=(c_1,c_2,c_3,c_4,c_5)^T$ and $d=(d_1,d_2,d_3,d_4,d_5)^T$ are pentaplets;  $f=(f_1,f_2,f_3,f_4)^T$ and $g=(g_1,g_2,g_3,g_4)^T$ are tetraplets.

\begin{itemize}
 \item $ [\mbf{3} \otimes \mbf{3}]_\mbf{3}  = i \Big(a_2 b_3 - a_3 b_2,~a_1 b_2 - a_2 b_1,~a_3 b_1-a_1 b_3\Big)^T$
 
 \item $[\mbf{3'} \otimes \mbf{3'}]_\mbf{3'} = i\Big(a'_2 b'_3 - a'_3 b'_2,~a'_1 b'_2 - a'_2 b'_1,~a'_3 b'_1-a'_1 b'_3\Big)^T$ 

 \item $[\mbf{3} \otimes \mbf{3'}]_\mbf{4} =  i\Big(a_2 b'_1 - \frac{a_3 b'_2}{\sqrt{2}},-a_1 b'_2 + \frac{a_3 b'_3}{\sqrt{2}},a_1 b'_3 - \frac{a_2 b'_2}{\sqrt{2}},-a_3 b'_1 + \frac{a_2 b'_3}{\sqrt{2}}\Big)^T$ 
 
 \item $[\mbf{3} \otimes \mbf{4}]_\mbf{3'}= i\Big(a_2 g_4 -a_3 g_1,\frac{1}{\sqrt{2}}(\sqrt{2} a_1 g_2 +a_2 g_1 +a_3 g_3),-\frac{1}{\sqrt{2}}(\sqrt{2} a_1 g_3 +a_2 g_2 +a_3 g_4)\Big)^T$ 
  
 \item $[\mbf{3} \otimes \mbf{4}]_{\mbf 4}= i\Big(a_1 g_1 + \sqrt{2} a_3 g_2,-a_1 g_2 + \sqrt{2} a_2 g_1,a_1 g_3 - \sqrt{2} a_3 g_4, -a_1 g_4 - \sqrt{2} a_2 g_3\Big)^T$

 \item $[\mbf{3}' \otimes \mbf{4}]_ \mbf{3}= i\Big(a'_2 g_3 -a'_3 g_2,\frac{1}{\sqrt{2}}(\sqrt{2} a'_1 g_1 +a'_2 g_4-a'_3 g_3),\frac{1}{\sqrt{2}}(-\sqrt{2} a'_1 g_4 +a'_2 g_2 -a'_3 g_1)\Big)^T$ 
 \item $[\mbf{3}' \otimes \mbf{4}]_ \mbf{4}= i\Big(a'_1 g_1 + \sqrt{2} a'_3 g_3,a'_1 g_2 - \sqrt{2} a'_3 g_4,-a'_1 g_3 + \sqrt{2} a'_2 g_1,
	      -a'_1 g_4 - \sqrt{2} a'_2 g_2\Big)^T$

 \item $[\mbf{3}' \otimes \mbf{5} ]_\mbf{5} = i\Big(a'_2 c_4 -a'_3 c_3,\frac{2 a'_1 c_2 +\sqrt{2}a'_3 c_4}{\sqrt{3}},-a'_2 c_1 -\frac{a'_1 c_3 -\sqrt{2}a'_3 c_5}{\sqrt{3}}  a'_3 c_1 +\frac{a'_1 c_4 +\sqrt{2}a'_2 c_2}{\sqrt{3}},\frac{-2 a'_1 c_5 +\sqrt{2}a'_2 c_3}{\sqrt{3}}\Big)^T
  $ 
 
 \item $[\mbf{4} \otimes \mbf{4}]_{\mbf{3}_a} =i\Big(f_1 g_4 -f_4 g_1 +f_3 g_2- f_2 g_3,\sqrt{2}(f_2 g_4 -f_4 g_2), \sqrt{2}(f_1 g_3 -f_3 g_1)\Big)^T $ 
 \item $[\mbf{4} \otimes \mbf{4}]_{\mbf{3'}_a} =i\Big(f_1 g_4 -f_4 g_1 +f_2 g_3- f_3 g_2,\sqrt{2}(f_3 g_4 -f_4 g_3), \sqrt{2}(f_1 g_2 -f_2 g_1)\Big)^T$ 
  \item $[\mbf{4} \otimes \mbf{4}]_{\mbf{4}_a} =i\Big(f_3 g_3 -f_4 g_2 -f_2 g_4,f_1 g_1 +f_3 g_4 +f_4 g_3, -f_4 g_4 -f_1 g_2 -f_2 g_1-f_2 g_2 +f_1 g_3 +f_3 g_1\Big)^T$ 
 
 \item $[\mbf{4} \otimes \mbf{5}]_{\mbf{5}_1} = 
 i\Big(f_1 c_5 +2 f_2 c_4 -2 f_3 c_3+f_4 c_2,-2 f_1 c_1+\sqrt{6} f_2 c_5 ,f_2 c_1+\sqrt{\frac{3}{2}}(-f_1 c_2 -f_3 c_5+2 f_4 c_4)-f_3 c_1-\sqrt{\frac{3}{2}}(f_2 c_2 +f_4 c_5+2 f_1 c_3),-2 f_4 c_1-\sqrt{6} f_3 c_2\Big)^T$ 
 \item $[\mbf{4} \otimes \mbf{5}]_{\mbf{5}_2} = 
   i\Big(f_2 c_4 - f_3 c_3,-f_1 c_1+\frac{2 f_2 c_5-f_3 c_4 -f_4 c_3}{\sqrt{6}},-\sqrt{\frac{2}{3}}(f_1 c_2 +f_3 c_5- f_4 c_4)
   -\sqrt{\frac{2}{3}}(f_1 c_3 +f_2 c_2+ f_4 c_5),-f_4 c_1-\frac{2 f_3 c_2+f_1 c_4 +f_2 c_3}{\sqrt{6}}\Big)^T$ 

 \item $[\mbf{5} \otimes \mbf{5}]_{\mbf{3}_a} =
 i\Big(2 ( c_4 d_3 -c_3 d_4)+c_2 d_5 -c_5 d_2,\sqrt{3}(c_2 d_1-c_1 d_2)+\sqrt{2}(c_3 d_5 -c_5 d_3)\sqrt{3}(c_5 d_1-c_1 d_5)+\sqrt{2}(c_4 d_2 -c_2 d_4)\Big)^T$ 
 \item $[\mbf{5} \otimes \mbf{5}]_{\mbf{3'}_a} =
 i\Big(2 ( c_2 d_5 -c_5 d_2)+c_3 d_4 -c_4 d_3,\sqrt{3}(c_3 d_1-c_1 d_3)+\sqrt{2}(c_4 d_5 -c_5 d_4)\sqrt{3}(c_1 d_4-c_4 d_1)+\sqrt{2}(c_3 d_2 -c_2 d_3)\Big)^T$
 \item $[\mbf{5} \otimes \mbf{5}]_{\mbf{4}_a} =
i \Big((c_1 d_2 -c_2 d_1)+\sqrt{\frac{3}{2}}(c_3 d_5-c_5 d_3),(c_1 d_3 - c_3 d_1)+\sqrt{\frac{3}{2}}(c_4 d_5-c_5 d_4)
(c_4 d_1 -c_1 d_4)+\sqrt{\frac{3}{2}}(c_3 d_2-c_2 d_3),(c_1 d_5 - c_5 d_1)+\sqrt{\frac{3}{2}}(c_4 d_2-c_2 d_4)\Big)^T   $
 \item $[\mbf{5} \otimes \mbf{5}]_{\mbf{4}_s} =
 i \Big((c_1 d_2 -c_2 d_1)+\sqrt{\frac{3}{2}}(c_3 d_5-c_5 d_3),(c_1 d_3 - c_3 d_1)+\sqrt{\frac{3}{2}}(c_4 d_5-c_5 d_4)
 (c_4 d_1 -c_1 d_4)+\sqrt{\frac{3}{2}}(c_3 d_2-c_2 d_3),(c_1 d_5 - c_5 d_1)+\sqrt{\frac{3}{2}}(c_4 d_2-c_2 d_4)\Big)^T $
\end{itemize}

%
\section{PMNS parametrization} \label{sec:Invariant}
%
We use the following convention for the PMNS matrix
\begin{align}
\label{PMNS_definition_general}
 \UPMNS = \tilde{U} \ {\rm diag}\{1, e^{i \alpha/2}, e^{i (\beta/2 + \delta)}\}
\end{align}
and $\tilde{U}$ is the CKM-like parametrization of the mixing matrix, defined as
\begin{align}
 \tilde{U} = \begin{pmatrix}
        1 & 0 & 0 \\
        0 & c_{23} & s_{23} \\
        0 & -s_{23} & c_{23} \\
       \end{pmatrix}
       \begin{pmatrix}
        c_{13} & 0 & s_{13} e^{-i \delta} \\
        0 & 1 & 0 \\
        -s_{13} e^{i \delta} & 0 & c_{13} \\
       \end{pmatrix}
	\begin{pmatrix}
        c_{12} & s_{12} & 0 \\
        -s_{12} & c_{12} & 0 \\
        0 &  0 & 1 \\
       \end{pmatrix}\ 
\end{align}
where $c_{\ij} \equiv \cos\theta_{ij}$ and $s_{ij} \equiv \sin\theta_{ij}$.
All the angles are in the first quadrant $\theta_{ij} \in [0, \pi/2]$. Here $\delta$ is the Dirac $CP$ phase, $\alpha$ and $\beta$ are the Majorana phases.
The mixing angles can be extracted using the PMNS matrix element as
 \begin{align}
 \label{mixing_angles_from_PMNS}
  \sin^2\theta_{12} = \frac{|U_{12}|^2}{1-|U_{13}|^2} \quad \sin^2\theta_{13} = |U_{13}|^2 \quad\sin^2\theta_{23} = \frac{|U_{23}|^2}{1-|U_{13}|^2}
 \end{align}
With this convention we can define the $\JCP$ invariant as
\begin{align}
\label{JCP_definition}
 \JCP \equiv  {\rm Im}{\bigg[U_{11} U^*_{13}U^*_{31}U_{33}\bigg]} = \frac{1}{8} \sin 2\theta_{12} \sin 2\theta_{23} \sin 2\theta_{13}\cos\theta_{13} \sin\delta.
\end{align}
The Dirac $CP$ phase can be extracted from \eqref{JCP_definition} and the mixing angles as
\begin{align} \label{eqn:deltaDirac}
 \sin \delta = \frac{8\JCP}{\sin 2\theta_{12} \sin 2\theta_{23} \sin 2\theta_{13}\cos\theta_{13}}.
\end{align}
Notice that the Dirac phase $\delta$ has a physical meaning only if all mixing angles are different from $0$
and $\pi/2$. \\
The Majorana phases can be extracted from the numerical PMNS mixing matrix taking into account the unphysical phases described by the diagonal matrix ${\rm diag}\{\exp{i \delta_e}, \exp{i \delta_\mu}, \exp{i \delta_\tau}\}$ that multiplies the $\UPMNS$ from the left.
Those can be eliminated with a redefinition of the charged lepton fields.
A similar procedure is discussed in \cite{Antusch:2003kp} using a different parametrization for the PMNS matrix.
We can obtain the Majorana phases as
\begin{align}
 \alpha &= 2 \arg \left\{\frac{U_{12}}{U_{11}} \right\}\quad \beta = 2 \arg \left\{\frac{U_{13}}{U_{11}}\right\}.
\end{align}
For sake of completeness we report the values of the unphysical phases
\begin{align}
 \delta_e = \arg\{U_{11}\}\quad 
 \delta_\mu = \arg \left\{U_{23}e^{-i (\beta/2 + \delta) } \right\}\quad
 \delta_\tau =\arg \left\{U_{33}e^{-i (\beta/2 + \delta) }\right\}.
\end{align}

%
\section{Equivalence among mechanisms}
\label{App:equivalence}
%
\begin{figure}
	\centering
	\includegraphics[scale=0.35]{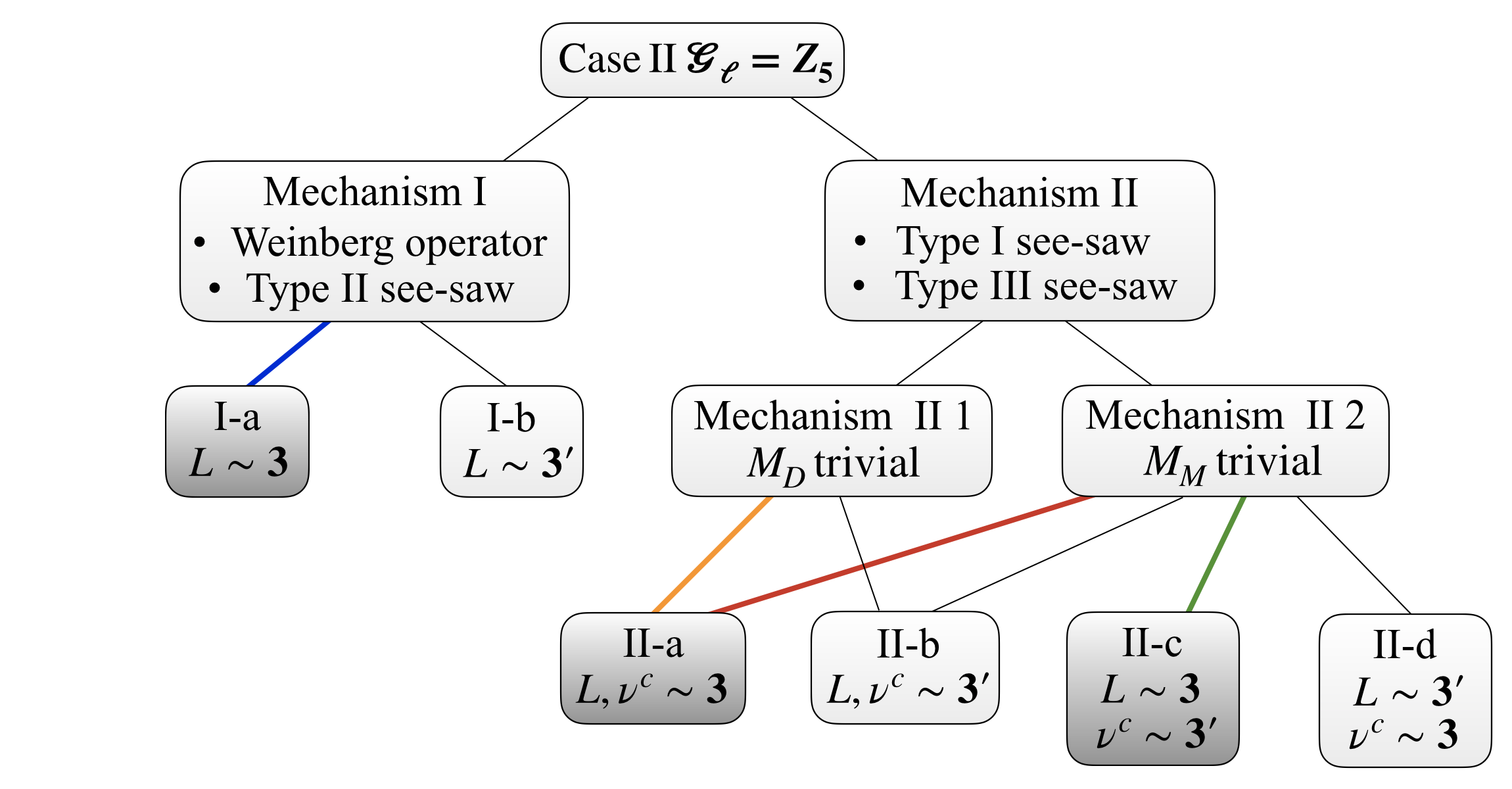}
	\caption{\label{fig:diagram_mechanism} \it \small Classification of all possible mechanisms to generate neutrino masses.
	The selected independent cases analysed in this article are highlighted with a dark gray background.}
\end{figure}
Here we show that some of the cases collected in Figure \ref{fig:diagram_mechanism} turn out to be equivalent to others in the sense that their phenomenology is the same under a redefinition of the model parameters.\\

{\bf $\bullet$ Weinberg operator and type II see-saw.} 
Neglecting flavour indices, the Weinberg operator may be written as
	\beq {\cal O}_{\rm Wein} \;=\; \frac{y_{\rm w}}{\Lambda}\, (L^Ti\sigma_2H)C(H^Ti\sigma_2L) \label{eqn:App_WB} 	
	\eeq
with $\Lambda$ the UV scale, $C$ the charge conjugation operator, $\sigma_2$ the Pauli matrix, $L$ the lepton doublet under $SU(2)_L$ and $H$ the Higgs doublet.
For type II see-saw models, the relevant Lagrangian is
	\beq {\cal L}_{\rm II} \;=\; y_{\rm II} L^TC\sigma_2\Delta L \:+\: {\rm h.c.} \label{eqn:App_II} \eeq
where $\Delta$ is a triplet scalar under $SU(2)_L$ which gets an induced vev, {\small $\langle\Delta\rangle=\mu\langle H\rangle^2/2m_\Delta^2$}, through the scalar potential
	\beq V \;=\; \mu H^T\sigma_2\Delta^\star H \:+\: m_\Delta^2\, {\rm Tr}[\Delta\Delta] \:+\: \cdots \eeq
when $\mu\neq 0$.
Inserting the vev in Eq.~\eqref{eqn:App_II} and comparing to Eq.~\eqref{eqn:App_WB}, one may see that the correspondent neutrino mass matrices are simply related by the parameter redefinition
	\beq\frac{y_{\rm w}}{\Lambda} ~\longleftrightarrow~ -y_{\rm II}\,\frac{\mu}{2m_\Delta^2} \eeq

{\bf $\bullet$ Type I and type III see-saw.} 
A similar correspondance can be established for type I and type III see-saw realizations.
For type I see-saw, the Lagrangian responsible for Majorana neutrino masses consist of the Dirac and Majorana terms
	\beq {\cal L}_{\rm I} \;=\; -Y_D\, \bar L \sigma_2 H^\star \nu^c \:+\: \frac{1}{2} M_{\rm R}\,\nu^c\nu^c \:+\: {\rm h.c.} \label{eqn:App_I} \eeq
where $\nu^c$ is the right-handed neutrino.
After EWSB, the light neutrino masses are generated as
	\beq M_\nu \;=\; -\langle H\rangle^2\, Y_D^T\, M_R^{-1} Y_D. \eeq
In type III see-saw models, the additional field is a fermion triplet under $SU(2)_L$, $\overrightarrow{T}$. The relevant Lagrangian for neutrino masses is
	\beq {\cal L}_{\rm I} \;=\; -Y_{\rm III}\, L^T C\sigma_2\, \overrightarrow{\sigma}.\overrightarrow{T}\,H \:+\: M_{\rm T} \overrightarrow{T}.\overrightarrow{T} \label{eqn:App_III} \eeq
As in the previous mechanism, the induced neutrino masses are
	\beq M_\nu \;=\; \langle H \rangle^2 Y_{\rm III}^T\, M_{\rm T}^{-1}\, Y_{\rm III}. \eeq
Since we are interested in correlations about the mass spectrum and mixing angles, we can investigate only the type I case without loss of generality.\\

{\bf $\bullet$ Mechanism I-a and I-b.}
Using the $A_5$ generators in representation ${\bf 3}'$, the PMNS of case II can be obtained with the representative touple $(Q, S, X) = (T, S, X_0)$.
The vev of the flavon in the five dimensional representation computed as in Section \ref{subsec:flavon_vevs} is then:  
	$$\langle  \phi_{\nu, {\bf 5}} \rangle^T \;=\; \left(-\sqrt{\dfrac{2}{3}}(x_r + x_{r,2}),\; -x_r + i(1- 
	\varphi) x_i,\; x_{r, 2} - i x_i,\; x_{r, 2} + i x_i,\;x_r + i(1- \varphi) x_i \right) $$
The same phenomenology is recovered under the following redefinition of vevs:
\begin{equation} \label{vev_redefinition_3_3prime}
	v_1 \longrightarrow v_1 \qquad x_r \longleftrightarrow x_{r,2}\qquad x_i \longrightarrow \varphi x_i\,.
\end{equation}
which is equivalent to a redefinition of the neutrino mass matrix parameters as
\begin{equation}
	s \longrightarrow s \qquad x \longleftrightarrow z \qquad y \longrightarrow \varphi y\,.
\end{equation}

{\bf $\bullet$ Mechanism II-a and II-b.}
Here there are two possibilities, a trivial $M_D$ or trivial $M_M$.
For a trivial $M_D$, one may check that the redefinition of the vevs is like in the previous case, but for the Majorana parameters (in capital letters): $S \longrightarrow S$, $X \longleftrightarrow Z$ and $Y \longrightarrow \varphi Y$.
In the second case, we have to consider the vev of the flavon in representation ${\bf 3}'$ invariant under $S=Z$, which is
	\beq \langle  \phi_{\nu, {\bf 3}'} \rangle^T \;=\; w \left(-\sqrt{2} \varphi^{-1},\, 1,\, 1\right) \eeq
Thus, with respect to the case $L, \nu^c \sim {\bf 3} \in A_5$, the redefinition of the vevs to generate the same mass matrix would be
\begin{align} \label{relations_vevs_model_E_and_F}
	v \longrightarrow w \qquad x_i \longrightarrow \varphi^{-1} x_i \qquad x_r \longleftrightarrow x_{r,2}
\end{align}

{\bf $\bullet$ Mechanism II-c and II-d.}
In this case we need to know the vev invariant under $Z=S$ for the quadruplet
	\beq \langle  \phi_{\nu, {\bf 4}} \rangle^T \;=\; (y_r-i y_i,\, (3-2\varphi)y_r -i y_i,\, (3-2\varphi)y_r +i 
	y_i,\, y_r+i y_i ) \eeq
If we consider the light neutrino mass matrix $M_\nu$ we observe that it is invariant under the vevs redefinition
\begin{align}
	x_r \longleftrightarrow \pm x_{r,2} \qquad x_i \longrightarrow \pm \varphi x_i \qquad y_r \longrightarrow \pm 
	(1 +2 \varphi) y_r \qquad y_i \longrightarrow \mp y_i.
\end{align}

%
\section{Definition of the $\chi^2$ function} \label{sec:chisquare}
%
We report in Table \ref{tab:Maltoni_et_al} the best fit points with the 1$\sigma$ errors and the 3$\sigma$ confidence region for NO and IO. All the data are extracted from the website \href{http://www.nu-fit.org}{\texttt{http://www.nu-fit.org}} \cite{Esteban:2016qun}. 
\begin{table}[H]
\begin{center}
\begin{tabular}{c  c  c  c c}
\toprule
\toprule
& \multicolumn{2}{c}{\bf Normal Ordering} & \multicolumn{2}{c}{\bf Inverted Ordering}\\
\midrule
\bf Parameter &  \bf Best Fit & \bf 3$\sigma$ Range &  \bf Best Fit & \bf 3$\sigma$ Range\\ 
\midrule
$\sin^2\ths/10^{-1}$ & 3.07$^{+0.13}_{-0.12}$ & 2.72 $\div$ 3.46 & 3.07$^{+0.13}_{-0.12}$ & 2.72 $\div$ 3.46\\
$\sin^2\tha/10^{-1}$ & 5.38$^{+0.33}_{-0.69}$ & 4.18 $\div$ 6.13 & 5.54$^{+0.23}_{-0.33}$ & 4.35 $\div$ 6.16\\
$\sin^2\thr/10^{-2}$ & 2.206$^{+0.075}_{-0.075}$ & 1.981 $\div$ 2.436 & 2.227$^{+0.074}_{-0.074}$ & 2.006 $\div$ 2.452 \\
$\delta/º$			 & 234$^{+43}_{-20}$  & 144 $\div$ 374 & 278$^{+26}_{-29}$ & 192 $\div$ 354 \\
$\Delta m^2_{21}/10^{-5}$ & 7.40$^{+0.21}_{-0.20}$ & 6.80 $\div$ 8.02 & 7.40$^{+0.21}_{-0.20}$ & 6.80 $\div$ 8.02\\
$\Delta m^2_{3\ell}/10^{-3}$ & +2.494$^{+0.033}_{-0.031}$ & +2.399 $\div$ +2.593 & -2.465$^{+0.032}_{-0.031}$ & -2.562 $\div$ -2.369 \\
\bottomrule
\bottomrule
\end{tabular}
\captionsetup{width=.9\textwidth}
\caption{\label{tab:Maltoni_et_al} \it \small Results for the global NuFIT 3.2 (2018).
Notice that in the last line for NO $\ell =1$ and for IO $\ell = 2$.
The analysis prefers a global minimum for NO, with $\Delta\chisquare = 4.7$ respect to the local minimum of IO.}
\end{center}
\end{table}
\vspace{-0.75cm} The admitted intervals of the absolute values of the elements of the PMNS mixing matrix at the 3$\sigma$ level are
\begin{align}
\label{PMNS_3_sigma_range}
 \|\UPMNS \| = \begin{pmatrix}
              0.799 \div 0.844 & 0.516 \div 0.582 & 0.141 \div 0.156 \\
              0.242 \div 0.494 & 0.467 \div 0.678 & 0.639 \div 0.774 \\
              0.284 \div 0.521 & 0.490 \div 0.695 & 0.615 \div 0.754
             \end{pmatrix}.
\end{align}
This result does not assume any particular neutrino mass ordering.
The ratio $r_\ell$ is defined as
\begin{align}
\label{r_ell_definition}
 r_\ell \equiv \frac{\Delta m^2_{21}}{\Delta m^2_{3\ell}} = \left\{\begin{array}{l l}
                                                             +2.967 \times 10^{-2} & \text{for NO} \\
                                                             -3.002 \times 10^{-2} & \text{for IO} 
                                                            \end{array} \right.
\end{align}
where we have used the best fit values. At the level of 3$\sigma$ the absolute value of $r_\ell$, defined in Eq. \eqref{r_ell_definition}, is constrained in the interval
\begin{align}
\label{Rnu_range_NO}
2.62 \times 10^{-2} \leq r_1 \leq 3.34 \times 10^{-2} 
\end{align}
for NO, while for IO
\begin{align}
\label{Rnu_range_IO}
-3.39 \times 10^{-2} \leq r_2 \leq -2.65 \times 10^{-2} .
\end{align}

We use the public data available at \href{http://www.nu-fit.org}{\texttt{http://www.nu-fit.org}} for the one dimensional projections $\chisquare_q$ of $q$-th parameter to construct the test function for the Standard Model point of the parameter space $\mathscr{Q}_\ell = \{\sin^2\theta_{12}, \sin^2{\theta_{13}}, \sin^2{\theta_{23}}, \Delta m^2_{21}, \Delta m^2_{3\ell}\}$
\begin{align}
	\chisquare({\bf q}) = \sum_{q \in \mathscr{Q}_\ell} \chisquare_q({\bf q}), 
\end{align}
where the sum is over all the observables assuming an ordering of the mass spectrum.\\
Since the available data are a discrete collection of points we use a first order polynomial function to interpolate the dataset.
In this way we are able to evaluate the $\chisquare_q$ for each point of the parameters space.
To quantify the contribution of the mixing angles and the mass splittings in the $\chisquare$ we introduce in our discussion the parameters $\chisquare_a$ and $\chisquare_m$ which are defined as
\begin{align}
 \chisquare_a \equiv \sum_{i\not= j} \chisquare_{ij}({\bf q}) \qquad \chisquare_m \equiv  \chisquare_{\Delta m_{21}^2}({\bf q}) +\chisquare_{\Delta m_{3\ell}^2}({\bf q})
\end{align}
where in $\chisquare_m$ for NO $\ell = 1$, and for IO $\ell = 2$.
The results are reported in Tables \ref{tab:mechanismI_chisquare} to \ref{tab:mechanismIIc2_chisquare}.
The general trend for every mechanism is a difficulty in the fit of the mixing  angles, whereas for the mass differences we observe a very low $\chi^2$, except for those cases where the global fit is bad.
This can be traced back to the problem in shifting the atmospheric angle from maximal value to the best fit one in Table \ref{tab:Maltoni_et_al}.

\begin{table}[h!]
\begin{center}
\begin{tabular}{c c c c c}
\toprule
\toprule
\bf  			&  $z=0$ 	& $x=0$		& $s=0$ - NO	& $s=0$ - IO $\xmark$\\
\midrule
$\chisquare_a$  &  $6.37$	& $6.25$	& $4.42$		&  $6.24$\\
$\chisquare_m$  &  $0.12$	& $0.06$	& $0.02$		&  $0.11$\\
\midrule
$\chisquare$  	&  $6.49$	& $6.31$	& $4.44$		&  $6.35$\\
\bottomrule
\bottomrule
\end{tabular}
\captionsetup{width=0.9\textwidth}
\caption{\it \small Minimum of the $\chisquare$ in the case of Mechanism I.
Analytic expressions as a series in the small expansion parameter could not be found for $s=0$ (IO).
Realizations $z=0$ and $x=0$ does not satisfy the latest combined bound from \texttt{\normalsize Planck} $\oplus$ BAO
on the sum of the neutrino masses, but only the \texttt{\normalsize Planck} limit \cite{Aghanim:2018eyx}.}
\label{tab:mechanismI_chisquare}
\end{center}
\end{table}

\begin{table}[h!]
\begin{center}
\begin{tabular}{c c c c c c  c c c}
\toprule
\toprule
\bf  		&  \multicolumn{2}{c}{$Z = 0$ (NO)} 	& \multicolumn{3}{c}{$X = 0$ (NO)} 	 		 & \multicolumn{2}{c}{$S = 0$ }\\
\bf  		&  $|S| = |X|$	& $|S|\gg|X|$	& $S \simeq Z/3$& $S \simeq-Z/4$& $X\simeq2Z/3$ & NO $\xmark$& IO $\xmark$\\
\midrule
$\chisquare_a$  & $4.37$	& $4.58$		& $4.47$		& $4.74$		& $4.41$		& $4.38$	& $6.84$\\
$\chisquare_m$  & $0.01$	& $0.07$		& $0.01$		& $0.02$		& $0.28$		& $0.04$	& $0.21$\\
\midrule
$\chisquare$  	& $4.38$	& $4.65$		& $4.48$		& $4.76$		& $4.69$		& $4.42$	& $7.05$\\
\bottomrule
\bottomrule
\end{tabular}
\captionsetup{width=0.9\textwidth}
\caption{\small \it Minimum of the $\chisquare$ in the case of Mechanism II a-1.
The cases ${\scriptsize Z=0 \wedge S\ll X}$ and ${\scriptsize S=0}$ (IO) are not compatible with the latest combined bound from \texttt{\normalsize Planck} $\oplus$ BAO: $\sum m_j=0.12$ eV @$95\%$ CL but only with $\sum m_j = 0.59$ eV @$95\%$ CL (\texttt{\normalsize Planck}, 2015) \cite{Ade:2015xua} and $\sum m_j=0.26$ eV @$95\%$ CL(\texttt{\normalsize Planck}, 2018) \cite{Aghanim:2018eyx}, respectively.}
\label{tab:mechanismIIa1_chisquare}
\end{center}
\end{table}

\begin{table}[h!]
\begin{center}
\begin{tabular}{c c c c c c  c c c}
\toprule
\toprule
\bf  		&  \multicolumn{2}{c}{$h_i = f = 0$} 	& {$h_i = h_r = 0$} 	& \multicolumn{2}{c}{$h_i = h_{r,2} = 0$}\\
\bf  			&  NO		&  IO	 &  NO		& NO $\xmark$ & IO \\
\midrule
$\chisquare_a$  & $4.67$	& $6.26$ & $11.15$	&  $4.36$ & $6.24$\\
$\chisquare_m$  & $0.15$	& $0.05$ &  $0.65$	&  $0.16$ & $0.36$\\
\midrule
$\chisquare$  	& $4.82$	& $6.31$ & $11.80$	&  $4.52$ & $6.60$\\
\bottomrule
\toprule
\bf  		&  \multicolumn{2}{c}{$g = f = 0$} 	& {$g = h_r = 0$} 		 & {$g = h_{r,2} = 0$}\\
\bf  		&		NO		&  IO		& NO		& IO\\
\midrule
$\chisquare_a$  & $4.60$	& $6.30$	& $10.08$	& $6.44$\\
$\chisquare_m$  & $0.08$	& $0.06$	&  $0.87$	& $0.12$\\
\midrule
$\chisquare$  	& $4.68$	& $6.36$	& $10.96$	& $6.56$\\
\bottomrule
\bottomrule
\end{tabular}
\captionsetup{width=0.9\textwidth}
\caption{\it Minimum of the $\chisquare$ in the case of Mechanism II a-2.
The cases associated with $h_r=0$ are not compatible with the latest combined bound from \texttt{\normalsize Planck} $\oplus$ BAO but only with the \texttt{\normalsize Planck} limit \cite{Aghanim:2018eyx}.}
\label{tab:mechanismIIa2_chisquare}
\end{center}
\end{table}

\begin{table}
\begin{center}
\begin{tabular}{c c c c c c  c c c}
\toprule
\toprule
\bf  		&  \multicolumn{2}{c}{$h_i = f_i = 0$} 	& {$h_i = h_r = 0$} 	& \multicolumn{2}{c}{$h_i = h_{r,2} = 0$}\\
\bf  		&  NO		&  IO	 	&  NO		& NO & IO $\xmark$\\
\midrule
$\chisquare_a$  & $4.37$	&  $6.32$	&  $8.23$	&   $6.21$	& $6.25$\\
$\chisquare_m$  & $0.01$	&  $0.08$	& $12.41$	&  $19.91$	& $0.05$\\
\midrule
$\chisquare$  	& $4.38$	&  $6.40$	& $20.64$	&  $26.12$	& $6.30$\\
\bottomrule
\toprule
\bf  			&  \multicolumn{2}{c}{$f_r = f_i = 0$} 	& {$f_r = h_r = 0$} 		 & {$f_r = h_{r,2} = 0$}\\
\bf  			&  NO		&  IO	 	& NO		& NO\\
\midrule
$\chisquare_a$  & $4.44$	&  $6.25$	& $4.37$	& $4.37$\\
$\chisquare_m$  & $0.12$	&  $0.06$	& $0.13$	& $0.01$\\
\midrule
$\chisquare$  	& $4.56$	&  $6.31$	& $4.50$	& $4.38$\\
\bottomrule
\bottomrule
\end{tabular}
\captionsetup{width=0.9\textwidth}
\caption{\it Minimum of the $\chisquare$ in the case of Mechanism II c-2.
The cases {\small $h_i=\lbrace h_r,\, h_{r,2}\rbrace=0$} (NO) get a substantially worse fit compared to other cases due to the latest combined bound from \texttt{\normalsize Planck} $\oplus$ BAO.
The case {\small $h_i=h_r=0$} (IO) is not compatible with the \texttt{\normalsize Planck} $\oplus$ BAO bound, but only with the \texttt{\normalsize Planck} limit.}
\label{tab:mechanismIIc2_chisquare}
\end{center}
\end{table}

\section{Neutrino mass sum rules} \label{sec:mass_sum_rule}
The sum rule is a generic function of the complex masses, $\Sigma = \Sigma( \tilde{m}_1,  \tilde{m}_2,  \tilde{m}_3)$, that is equal to zero in a given model.
Using the same notation of \cite{Barry:2010yk}, the complex masses are defined as
	\begin{align}
		\tilde{m}_1 \equiv m_1 \qquad \tilde{m}_2 \equiv m_2 e^{i \alpha} \qquad \tilde{m}_3 \equiv m_3 e^{i 
		\beta},
	\end{align}
where $m_j$ are the absolute values of the light neutrino masses and $\alpha$ and $\beta$ are the Majorana phases, see \eqref{PMNS_definition_general} for our PMNS convention.
In particular, sum rules can be derived for our models.
In general the masses appear as $\tilde{m}_j^p$ where $p \in \mathbb{Z}$ depends on the type of neutrino masses.
For instance in the case of Mechanism I $p= 1$, in Mechanism II with trivial $m_D$ $p = -1$ and otherwise $p = 2$.

\newpage

%
\bibliographystyle{JHEP}
\bibliography{references.bib}
%

\end{document}